\documentclass[aps,prb,twocolumn]{revtex4}
\usepackage{graphicx}
\usepackage{epsfig}
\usepackage{graphics}
\usepackage{epsfig}
\usepackage{yfonts}
\usepackage{amsmath}
\usepackage{amsfonts}
\usepackage{amssymb}
\usepackage{xcolor,cancel}
\usepackage{bm}
\usepackage{float}
\usepackage{caption}
\usepackage{subcaption}
\usepackage[force]{feynmp-auto}

\newcommand{\braket}[1]{\ensuremath{\langle{#1}\rangle}}

\begin{document}

\title{Quantum transport in driven systems with vibrations: Floquet nonequilibrium Green’s functions and the GD approximation}

\author{Thomas D. Honeychurch} 
\author{Daniel S. Kosov}
\affiliation{College of Science and Engineering, James Cook University, Townsville, QLD, 4811, Australia}

\begin{abstract}
	We investigate the effects of alternating voltage on nonequilibrium quantum systems with localised phonon modes. Nonequilibrium Green's functions are utilised, with electron-phonon coupling being considered with the $GD$ approximation (self-consistent Born approximation). Using a Floquet approach, we assume periodicity of the dynamics. This approach allows us to investigate the influence of the driven electronic component on the nonequilibrium occupation of the vibrations. It was found that signatures of inelastic transport gained photon-assisted peaks. A simplistic model was proposed and found to be in good agreement with the full model in certain parameter ranges. Moreover, it was found that driving the alternating current at resonance with vibrational frequencies caused an increase in phonon occupation.
\end{abstract}

\maketitle

\section{Introduction}

The transport properties of quantum dots, especially molecular junctions, can be significantly altered by vibrations coupled to the central region, causing an array of interesting phenomena\cite{cuevas2010molecular}. Of particular importance is how vibrations, often in conjunction with other phenomena, inhibit device functionality and stability. 

Investigations that cast vibrations as localised phonons have been extensively studied under various parameter ranges. When coupling between electronic and vibrational components is sufficiently small, differential conductance through a junction has been found to vary due to changes in bias voltage, with electrons inelastically interacting with central phonons\cite{Galperin2004a,Galperin2004b,cuevas2010molecular}, explaining the experimental phenomena observed in inelastic electron tunnelling spectroscopy and point contact spectroscopy experiments\cite{djukic2005stretching,tal2008electron,you2017recent}. For sufficiently large coupling between electrons and phonons, transport through the system is suppressed due to the Franck-Condon blockade \cite{cuevas2010molecular,ryndyk2016theory}. With semi-classical approaches to mechanical change within a junction, electron-friction and non-adiabatic effects with general potentials can be studied\cite{preston2020current,Hedegard2010,Preston2021_first_passage,Bode2012,Lu2012,Kershaw2020}.

In steady state, vibrations within a molecular junction have been investigated with a variety of methods \cite{Galperin2007}. Nonequilibrium Green's functions approaches have been used to model vibrations within self-consistent perturbation theory\cite{Galperin2004a,Galperin2004b,Park2011,Dash2010}, polaron and dressed tunnelling approximations \cite{Maier2011,Souto2014}, equation of motion methods\cite{Galperin2006} and many more. Vibrations can also be studied with master equation approaches, like the Redfield master equations \cite{Peskin2020} or more involved methods like hierarchical quantum master equations\cite{Schinabeck2020}.

Vibrations often contribute significantly to junction failure. With the current lifetime of many molecular junctions being, at best, only seconds long \cite{Darwish2020}, tackling the problem a junction instability stands as a significant hurdle for the field\cite{Thoss2021,preston2020current}.

The further addition of time-dependence in the form of varying voltages and electric fields is frequent within the theory and experiment surrounding molecular electronics, allowing for the probing and control of dynamics within the junction. A prime example is recent work that has seen molecular junctions probed on picosecond time-frames with a laser pulse-pair scheme \cite{Arielly2017}.
Time-dependent potentials also allow for the realisation of novel functionalities, including time-dependent molecular rectifiers\cite{Tu2006,Trasobares2016} and molecular pumps\cite{Haughian2017}. Beyond static driving, molecular junctions are often studied within time-dependent settings, like transience\cite{Ridley2018,Ridley2022}, periodic driving of lead energies\cite{Honeychurch2020}, couplings\cite{Erpenbeck2019} or by laser pulses \cite{Beltako2019,Ochoa2015}, when the central junction is subject to monochromatic electric fields \cite{Cabra2020}, or within the limit of slow drivings \cite{Honeychurch2019,Thoss2022}.  

Given the importance of vibrations, understanding their dynamics under various time-dependent scenarios will be essentials for molecular junction designs that seek to capitalise on time-dependent effects. Of recent, exploration into the effects of time-dependent driving upon vibrations has been growing: transient dynamics of vibrationally active molecular junctions have been investigated theoretically with the self-consistent perturbation theory \cite{Souto2018} and dressed tunneling approximation \cite{Souto2015}; harmonic driving of a gate voltages was used to increase current within the Franck-Condon blockade region \cite{Haughian2016}; vibrations with perturbatively slow driving have been investigated with mean-field in a nonequilibrium Green's functions setting\cite{Honeychurch2019} and with Hierarchical master equations\cite{Thoss2022}; and Nonequilibrium Green's functions and linear response theory have been utilised to investigate conductance profiles and properties of phonon under small drivings\cite{Ueda2017,Ueda2016,Ueda2011}.  

Of particular interest is whether time-dependent driving can be used to reduce vibrations whilst still allowing for current flow comparable to equivalent static case. This has been predicted with a master equation approach\cite{Peskin2020}, and for vibrations modelled semi-classically with a Langevin approach\cite{Preston2020}. 

Within this paper, we use a Floquet nonequilibrium Green's function approach to investigate the effects of time-periodic driving on a single-level electronic molecule coupled to a single phonon mode, making use of the self-consistent perturbation theory in the form of the $GD$ approximation\cite{Souto2018,Karlsson2020,Ridley2022}.

It was found that changes in conductance, indicative of inelastic electron transport spectroscopy, gain photon-assisted side-peaks. Following intuition from photon-assisted transport and inelastic electron transport spectroscopy, a simplistic form for the current and phonon occupation is hypothesised and found to be a good match in limiting cases. It was also observed that resonances between the vibrational and driving frequencies resulted in increases to phonon occupation, with two different contributing mechanisms.

This paper is organized as follows: section II covers the theory. In section III, the method is applied to a single electronic level coupled with a phonon. In section IV, the results of the paper are summarized. Atomic units are used throughout the paper.

\section{Theory}

To describe electrons moving through a junction whilst interacting with central phonons, we make use of a nonequilibrium Green's functions approach, considering the electron-vibration coupling within the $GD$ approximation, also known as the self-consistent Born approximation.\cite{Schuler2016,Karlsson2020,Sakkinen2015}

\subsection{Hamiltonian}

The system is modelled with the following Hamiltonian,
\begin{equation}
H(t)=H_{el}(t) + H_{vib} + H_{e-v}.
\label{hamiltonian}
\end{equation}
The electronic component are given by:
\begin{equation}
\begin{split}
H_{el}(t) = H_{central} + H_{leads}(t) + H_{coupling},
\end{split}
\end{equation}
\begin{equation}
H_{central} (t) = \sum_{ij} \epsilon_{ij}  d^\dag_i d_{j},
\end{equation}
\begin{equation}
\label{leads}
H_{leads} (t)=  \sum_{k,\alpha=L,R}  \epsilon_{k\alpha}(t) c^{\dagger}_{k\alpha} c_{k\alpha},
\end{equation}
\begin{equation}
H_{coupling} (t) = \sum_{i k \alpha=L,R } t_{k\alpha i} \; c^{\dagger}_{k \alpha} d_i +\mbox{h.c.}
\end{equation}
and the Hamiltonian for the vibrations is given as 
\begin{equation}
H_{vibratons} = \sum_{\alpha} \omega_\alpha a^\dagger_\alpha a_\alpha,
\end{equation} 
and the coupling between the electronic and vibrational components is given by 
\begin{equation}
H_{e-v} = \sum_{ij, \alpha} \lambda^\alpha_{i,j} Q_{\alpha} d^\dagger_i d_j.
\end{equation}
Here, $d_i$ ($d^\dagger_i$) and $c_{k\alpha}$ ($c^\dagger_{k\alpha})$ are annihilation (creation) operators for the site $i$ in the central region and $k\alpha$ in the leads, respectively. For the phonons, the quantum operator for position is given by $Q_{\alpha} = \frac{1}{\sqrt{2}} \left(a_\alpha + a^\dagger_\alpha\right)$ and the momentum given by $P_\alpha = \frac{1}{\sqrt{2}i}\left(a_\alpha - a^\dagger_\alpha \right)$, where $a_\alpha$ $(a^\dagger_\alpha)$ is the annihilation (creation) operator for the phonon $\alpha$. Within the investigation, only the energies of the leads are considered to be time-dependent. It is a simple extension to consider time-dependence within the central energy levels, $\epsilon_{ij}$ and couplings, $t_{k\alpha,i}$.

\subsection{Nonequilibrium Green's functions}

To capture the dynamics of the system out of equilibrium, we make use of a nonequilibrium Green's functions approach \cite{Schuler2016,Haug2008,Stefanucci2013}. On the Keldysh contour, we have the electronic contour Green's function:
\begin{equation}
G_{ij} (\tau,\tau') = -i\left\langle T_c\left(d_i \left(\tau\right)d^\dagger_j \left(\tau'\right)\right) \right\rangle.
\end{equation}
Similarly, we have the phononic Green's functions,
\begin{equation}
\begin{split}
D_{\alpha \beta} (\tau,\tau') = -i\left\langle T_c\left(\Delta Q_\alpha \left(\tau\right)\Delta Q_\beta \left(\tau'\right)\right) \right\rangle
\\=
-i\left[\left\langle T_c\left(Q_\alpha \left(\tau\right) Q_\beta \left(\tau'\right)\right) \right\rangle - \left\langle Q_\alpha \left(\tau\right) \right\rangle  \left\langle Q_\beta \left(\tau'\right) \right\rangle  \right]
\end{split}
\end{equation}
and
\begin{equation}
\begin{split}
D^{PP}_{\alpha \beta} (\tau,\tau') = -i\left\langle T_c\left(\Delta P_\alpha \left(\tau\right)\Delta P_\beta \left(\tau'\right)\right) \right\rangle
\\=
-i\left[\left\langle T_c\left(P_\alpha \left(\tau\right) P_\beta \left(\tau'\right)\right) \right\rangle - \left\langle P_\alpha \left(\tau\right) \right\rangle  \left\langle P_\beta \left(\tau'\right) \right\rangle  \right],
\end{split}
\end{equation}
where $\Delta Q_\alpha \left(\tau\right) = Q_\alpha \left(\tau\right) - \left\langle Q_\alpha \left(\tau\right) \right\rangle$ and $\Delta P_\alpha \left(\tau\right) = P_\alpha \left(\tau\right) - \left\langle P_\alpha \left(\tau\right) \right\rangle$. The corresponding noninteracting Green's functions are denoted with lower case lettering. 

The current from the left or right lead is given by,
\begin{equation}
	\begin{split}\label{eq:current}
		I_{\alpha} (t) =  2 Re \left\{\int^{\infty}_{-\infty} dt_1 Tr\left[G^<(t,t_1) \Sigma_{\alpha}^A (t_1,t) \right.\right.
		\\
		\left.\left. + G^R (t,t_1) \Sigma_{\alpha}^< (t_1,t)\right]\right\},
	\end{split}
\end{equation}
and the occupation of the electrons within the central region is given by 
\begin{equation}\label{eq:electronic_occupation}
	n^{el}_{i} (t) = -i G_{ii}^< (t,t).
\end{equation}
For the phonons, we are principally interested in the phonon occupation:
\begin{equation}\label{eq:phonon_occupation}
	\begin{split}
		n^{ph}_{\alpha} (t) = \left\langle a^\dagger_{\alpha} \left(t\right)a_{\alpha} \left(t\right)\right\rangle
		\\
		= \frac{1}{2} \left[\braket{P_{\alpha} (t)^2} +\braket{Q_{\alpha} (t)^2}\right] - \frac{1}{2} 
		\\
		= \frac{1}{2} \left[ i D^<_{\alpha\alpha} \left(t,t\right) + \left\langle Q_{\alpha} (t) \right\rangle^2 \right. 
		\\
		\left. + i D^{<,PP}_{\alpha\alpha} \left(t,t\right) + \left\langle P_{\alpha} (t)\right\rangle^2\right] - \frac{1}{2}.
	\end{split}
\end{equation}

\subsection{Equations of motion}

To calculate the electronic and phononic nonequilibrium Green's functions, we use the Kadanoff-Baym equations:
\begin{equation}
\begin{split}
i\frac{\partial}{\partial \tau} G_{ij} \left(\tau,\tau'\right) - \sum_{k}  \epsilon_{ik} G_{kj} \left(\tau,\tau'\right) \\- \sum_{k} \int_c d\tau_1 \Sigma_{ik} \left(\tau,\tau_1\right) G_{kj} \left(\tau_1,\tau'\right) = \delta_{ij} \delta_c \left(\tau-\tau'\right)
\end{split}
\end{equation}
and
\begin{equation}
\begin{split}
\frac{-1}{\omega_\alpha} \left(\frac{d^2}{d\tau^2} +\omega_\alpha^2\right)  D_{\alpha\beta}\left(\tau,\tau'\right) = \delta_c \left(\tau-\tau'\right)\delta_{\alpha\beta} 
\\ + \sum_\gamma \int d\tau_1 \Pi_{\alpha\gamma}(\tau,\tau_1)D_{\gamma\beta}(\tau_1,\tau').
\end{split}
\end{equation}
Here, the external influences on the central electrons and phonons is captured in $\Sigma_{ij} (\tau,\tau')$ and $\Pi_{\alpha\beta} (\tau,\tau')$, respectively.
The collective influence on the central regions electrons can be separated into that due to the phonons and the leads:
\begin{equation}
\Sigma \left(\tau,\tau'\right) = \Sigma_{int} \left(\tau,\tau'\right) + \Sigma_{leads}\left(\tau,\tau'\right)
\end{equation}  
and, similarly for the phonons, 
\begin{equation}\label{eq:keldysh_phonon}
\Pi \left(\tau,\tau'\right) = \Pi_{int} \left(\tau,\tau'\right) + \Pi_{bath} \left(\tau,\tau'\right).
\end{equation}

Calculating the electronic lead self-energies follows the standard procedure: 
\begin{equation} \label{eq:self energy definition}
	\Sigma^{<,>,R,A}_{\alpha,ij} (t,t') = \sum_{k,k'} t^*_{k\alpha,i}(t)  \; g^{<,>,R,A}_{k\alpha,k'\alpha}\left(t,t' \right)  t_{k'\alpha,j}(t'),
\end{equation}
where the noninteracting lead self-energies follow the standard definitions,
\begin{equation}
	g_{k\alpha,k' \alpha'}^< (t,t') = i  f_{k\alpha} e^{-i \int_{t'}^{t} dt_1 \epsilon_{k\alpha} (t_1)} \delta_{k, k'},
\end{equation}
\begin{equation}
	g_{k\alpha,k'\alpha}^> (t,t') = -i (1-f_{k\alpha}) e^{-i \int_{t'}^{t} dt_1 \epsilon_{k\alpha} (t_1)} \delta_{k, k'},
\end{equation}
\begin{equation}
	g^R_{ka,k'a} (t,t') = -i \Theta \left(t-t'\right)  e^{-i \int_{t'}^{t} dt_1 \epsilon_{k\alpha} (t_1)} \delta_{k,k'},
\end{equation}
\begin{equation}
	g^A_{ka,k'a} (t,t') = i \Theta \left(t'-t\right)  e^{-i \int_{t'}^{t} dt_1 \epsilon_{k\alpha} (t_1)} \delta_{k,k'}.
\end{equation}	
The time-dependence takes the form $\epsilon_{k\alpha} (t) = \epsilon_{k\alpha} + \phi_\alpha (t)$, which allows us to separate out the phase induced by the varying energies of the leads from rest of the self-energy:
\begin{equation} \label{eq:lead_self_energies}
	\begin{split}
		\Sigma_{\alpha,ij} (t,t') 
		= \Sigma'_{\alpha,ij} (t - t') e^{-i \int_{t'}^{t} dt_1 \phi_{\alpha}(t_1)}
		\\=  e^{-i\Phi_{\alpha}(t)} \Sigma'_{\alpha,ij} (t - t') e^{i\Phi_{\alpha}(t')},
	\end{split}
\end{equation}
where $\Phi_{\alpha}(t)$ is the anti-derivative of $\phi_\alpha(t)$ and $\Sigma'_{\alpha,ij} (t - t')$ is the self-energy of the equivalent static case, which are taken in the wide-band approximation:
\begin{equation}
	\Sigma^{A/R}_{\alpha,ij}(\omega) = \pm\frac{i}{2} \Gamma_{\alpha,ij},
\end{equation}
\begin{equation}
	\Sigma^{<}_{\alpha,ij} (\omega) = i f_{\alpha} \left(\omega\right) \Gamma_{\alpha,ij},
\end{equation}
and
\begin{equation}
	\Sigma^{>}_{\alpha,ij} (\omega) = -i
	\left(1 - f_{\alpha} \left(\omega\right)\right)  \Gamma_{\alpha,ij},
\end{equation}
where the Fermi-Dirac occupation is given by:
\begin{equation}
	f_{k\alpha} = \frac{1}{1 + e^{(\epsilon_{k\alpha}-\mu_\alpha)/T_\alpha}}.
\end{equation}

Within the investigation, the lead energies were driven sinusoidally,
\begin{equation}\label{eq:phase_definition}
	\phi_{\alpha} (t) = \Delta_\alpha \cos (\Omega_\alpha t),
\end{equation}
which gives $\Phi(t)=\left(\Delta_\alpha / \Omega_\alpha \right)\sin \left(\Omega_\alpha t\right)$, which can be expressed as a Fourier series with the Jacobi-Anger expansion:
\begin{equation}\label{eq:jacobi_anger}
	e^{i \frac{\Delta_\alpha}{\Omega_\alpha}\sin(\Omega_\alpha t)} = \sum_{n=-\infty}^{n=\infty} J_n \left(\frac{\Delta_\alpha}{\Omega_\alpha}\right)  e^{in\Omega_\alpha t},
\end{equation} 
where $J_n(x)$ are Bessel functions of the first kind.

For the noninteracting phonons, we have the following phonon Green's functions:
\begin{equation}
	d_{\alpha\beta}^R (\omega) = \left[\frac{\frac{1}{2}}{\omega - \omega_\alpha + i\eta_\alpha} - \frac{\frac{1}{2}}{\omega + \omega_\alpha + i\eta_\alpha}\right]\delta_{\alpha\beta},
\end{equation}
\begin{equation}
	d_{\alpha\beta}^A (\omega) = \left[ \frac{\frac{1}{2}}{\omega - \omega_\alpha - i\eta_\alpha} - \frac{\frac{1}{2}}{\omega + \omega_\alpha - i\eta_\alpha}\right]\delta_{\alpha\beta},
\end{equation}
\begin{equation}\label{eqn:noninteracting_D_lesser}
	\begin{split}
		d_{\alpha\beta}^< (\omega) = - \pi i \left[f_B (\omega_\alpha) \delta (\omega - \omega_\alpha) \right.
		\\
		\left. + (1  + f_B (\omega_\alpha)) \delta (\omega + \omega_\alpha)\right]\delta_{\alpha\beta},
	\end{split}
\end{equation}
\begin{equation}\label{eqn:noninteracting_D_greater}
	\begin{split}
		d_{\alpha\beta}^> (\omega) = - \pi i \left[f_B (\omega_\alpha) \delta (\omega  + \omega_\alpha) \right.
		\\
		\left. + (1  + f_B (\omega_\alpha)) \delta (\omega - \omega_\alpha)\right]\delta_{\alpha\beta}.
	\end{split}
\end{equation}
Instead of letting $\eta_\alpha$ be infinitesimals, we take them as finite, so to capture the influence of a phonon bath on central phonons. Making use of fluctuation-dissipation relations, we can introduce $\eta_\alpha$ into the lesser and greater phonon Green's functions:

\begin{equation}
	\begin{split}
		d_{\alpha}^< (\omega) = \left(d_{\alpha}^R (\omega) - d_{\alpha}^A \left(\omega\right)\right) f_B (\omega)
		\\
		= \left(-\frac{i\eta_\alpha}{\left(\omega - \omega_\alpha\right)^2 + \eta_{\alpha}^2} + \frac{i\eta_\alpha}{\left(\omega + \omega_\alpha\right)^2 + \eta_{\alpha}^2}\right) f_B(\omega)
	\end{split}
\end{equation}
and
\begin{equation}
	\begin{split}
		d_\alpha^> (\omega) = \left(d_\alpha^R (\omega) - d_\alpha^A \left(\omega\right)\right)\left(1+f_B (\omega)\right)
		\\
		= \left(-\frac{i\eta_\alpha}{\left(\omega - \omega_\alpha\right)^2 + \eta_\alpha^2} + \frac{i\eta_\alpha}{\left(\omega + \omega_\alpha\right)^2 + \eta_\alpha^2}\right) \left(1+f_B (\omega)\right),
	\end{split}
\end{equation}
where taking the limit of $\eta$,
\begin{equation}
	\lim_{\eta \rightarrow 0^+} \left[\frac{\eta}{\left(\omega-\omega_0\right)^2 + \eta^2}\right] \rightarrow \pi \delta \left(\omega-\omega_0\right),
\end{equation}
and substituting for $f^0_B(-\omega) = - \left(f^0_B(\omega) + 1\right)$, gives us back the equations  (\ref{eqn:noninteracting_D_lesser}) and (\ref{eqn:noninteracting_D_greater}).

For the lesser and greater phonon Green's functions, we can use the fluctuation-dissipation rules to cast the effects due to the infinitesimals as self-energies. Equating equations  (\ref{eqn:noninteracting_D_lesser}) and (\ref{eqn:noninteracting_D_greater}) with the associated Keldysh equations gives us:
\begin{equation}
	\Pi_{\alpha}^{</>} \left(\omega\right) = \mp 4 i \eta_{\alpha} \left(\frac{\omega}{\omega_\alpha}\right) f_B (\pm\omega).
\end{equation}
This phonon self-energy is used to incorporate the effects of the infinitesimal into self-consistent calculations (see equation (\ref{eq:keldysh_phonon})).

In addition to the above Green's functions, we need to calculate $\left\langle Q_{\alpha} (\tau) \right\rangle$, $\left\langle P_{\alpha} (\tau)\right\rangle$ and $D^{PP}_{\alpha,\alpha'} \left(\tau,\tau'\right)$, allowing for the calculation of the phonon occupation. The equation of motion for the average position is given as \cite{Schuler2016},
\begin{equation} \label{eq:position_eom}
\begin{split}
\frac{-1}{\omega_\alpha} \left(\frac{d^2}{d\tau^2} +\omega_\alpha^2\right)\left\langle Q_{\alpha} (\tau) \right\rangle = \sum_{ij} -i \lambda^{\alpha}_{ij} G_{ji} \left(\tau,\tau^+\right) \\ +  \int d\tau_1 \Pi^{bath}_{\alpha}(\tau,\tau_1) \left\langle Q_{\alpha} (\tau_1) \right\rangle.
\end{split}
\end{equation}
For the terms with momentum operators, we make use of 
\begin{equation}
\frac{d}{d\tau} Q_\alpha (\tau) = \omega_\alpha P_\alpha (\tau),
\end{equation}
which gives us the equations of motions
\begin{equation}\label{eqn:eom_momentum}
\frac{d}{d\tau} \braket{Q_\alpha (\tau)} = \omega_\alpha \braket{P_\alpha (\tau)}
\end{equation}
and
\begin{equation}\label{eqn:eom_momentum_DF}
\begin{split}
\frac{d}{d\tau d\tau'} D_{\alpha \beta} \left(\tau,\tau'\right) = \omega_\alpha \delta_{\alpha\beta} \delta\left(\tau,\tau'\right)
\\
+ D^{PP}_{\alpha \beta} \left(\tau,\tau'\right) \omega_\alpha \omega_{\beta}.
\end{split}
\end{equation}
These objects allows us to calculate occupation:
\begin{equation}
\begin{split}
n^{ph}_{\alpha} (\tau) = \left\langle T_c\left(a_{\alpha} \left(\tau\right)a_{\alpha}^\dagger \left(\tau^+\right)\right) \right\rangle
\\= \frac{1}{2} \left[ \left\langle T_c\left(Q^2_{\alpha}\left(\tau\right)\right) \right\rangle + \left\langle T_c\left(P^2_{\alpha}\left(\tau\right)\right) \right\rangle\right] -\frac{1}{2}
\\
= \frac{1}{2} \left[ i D_{\alpha\alpha} \left(\tau,\tau^+\right) + \left\langle Q_{\alpha} (\tau) \right\rangle^2 \right. \\ \left. + i D^{PP}_{\alpha\alpha} \left(\tau,\tau^+\right) + \left\langle P_{\alpha} (\tau)\right\rangle^2 \right] - \frac{1}{2}.
\end{split}
\end{equation} 

The interaction between electrons and phonons can be approximated within the GD approximation  \cite{Sakkinen2015,Karlsson2020}:
\begin{equation}
\mathbf{\Sigma}_{ij}^{int} (\tau,\tau') = \mathbf{\Sigma}_{ij}^{har} (\tau,\tau') + \mathbf{\Sigma}_{ij}^{XC} (\tau,\tau'),
\end{equation} 
where
\begin{equation}\label{eqn:sigma_hartree}
\begin{split}
\mathbf{\Sigma}_{ij}^{har} (\tau,\tau') \\
= -i \delta \left(\tau,\tau'\right) \sum_{\beta}\lambda^\beta_{ij} \int_c d\tau_1 d_{\beta}(\tau,\tau_1) \sum_{ml} \lambda_{ml}^{\beta} G_{lm}(\tau_1,\tau_1^+),
\end{split}
\end{equation}
\begin{equation}\label{eqn:sigma_exchange}
\mathbf{\Sigma}^{XC}_{ij} (\tau,\tau') = i \sum_{\mu\nu, ml} D_{\mu\nu}\left(\tau,\tau'\right) \lambda^{\mu}_{im} G_{ml}\left(\tau,\tau'\right) \lambda^{\nu}_{lj},
\end{equation}
and
\begin{equation}\label{eqn:sigma_polarisation}
	\Pi^{int}_{\alpha\beta} (\tau,\tau') = -i \sum_{mlkp} \lambda^{\alpha}_{ml} G_{lk}\left(\tau,\tau'\right) \lambda^{\beta}_{kp} G_{pm}\left(\tau',\tau\right).
\end{equation}

\subsection{Floquet Theory}

Moving the equations of motion from the contour to real time with the greater, lesser, retarded and advanced projections \cite{Haug2008,Rammer2007}, we can solve the equations of motion with a Floquet approach \cite{Honeychurch2020,Brandes1997}, where we assume that the system is time-periodic around the central time $T=\frac{t+t'}{2}$, and complete a Fourier transform with respect to the relative time, $\tau = t-t'$: 
\begin{equation}\label{eq:periodicity_1}
A(t,t') = A(T,\tau) = \sum^{\infty}_{n= -\infty} A(\tau,n) e^{\Omega inT}
\end{equation}
and
\begin{equation}\label{eq:periodicity_2}
A\left(\omega,n\right) = \frac{1}{P} \int^P_{0} dT \; e^{-i\Omega n T} \int^{\infty}_{-\infty} d\tau e^{i\omega \tau} A(T,\tau).
\end{equation} 

For solving the convolutions of the form 
\begin{equation}
C(t,t') = \int dt_1 A(t,t_1) B(t_1,t'),
\end{equation}
we recast the Fourier coefficients into a Floquet matrix,
\begin{equation} \label{eqn:floquet_matrix}
\bar{A} \left(\omega,m,n\right) = A \left(\omega + \frac{\Omega}{2} \left(m + n\right), n-m\right), 
\end{equation}
which allows us to express the convolution as a matrix multiplication,
\begin{equation} \label{eq:floquet_convolution}
\bar{C} \left(\omega,m,n\right)  = \sum_{r=-\infty}^{\infty} \bar{A} \left(\omega,m,r\right) \bar{B} \left(\omega,r,n\right),
\end{equation}
allowing for the Kadanoff-Baym equations to be written as a matrix equation:
\begin{equation}
\begin{split}
\left(\omega + \Omega m \right)\bar{G}_{ij}^{R/A} \left(\omega,m,n\right) - \sum_{k} \epsilon_{ik} \bar{G}_{kj}^{R/A} \left(\omega,m,n\right) = 
\\
\delta_{ij}\delta_{mn} + \sum_{k,r} \bar{\Sigma}_{ik}^{R/A} \left(\omega,m,r\right) \bar{G}_{kj}^{R/A} \left(\omega,r,n\right),
\end{split}
\end{equation}
\begin{equation}
\begin{split}
\bar{G}_{ij}^{<} (\omega,m,n) =
\\
\sum_{kw,rs} \bar{G}_{ik}^R (\omega,m,r) \bar{\Sigma}^<_{kw} \left(\omega,r,s\right) \bar{G}_{wj}^A \left(\omega,s,n\right),
\end{split}
\end{equation}
\begin{equation}
\begin{split}
\frac{-1}{\omega_\alpha}\left(\omega^2_\alpha - \left(\omega + \Omega m \right)^2\right) \bar{D}^R_{\alpha\beta} \left(\omega,m,n\right)
=
\delta_{\alpha\beta}\delta_{mn} 
\\
+ \sum_{\gamma,r}  \bar{\Pi}_{\alpha\gamma} \left(\omega,m,r\right) \bar{D}_{\gamma\beta} \left(\omega,r,n\right),
\end{split}
\end{equation}
\begin{equation}
\begin{split}
\bar{D}_{\alpha\beta}^{<} (\omega,m,n)
\\ = \sum_{\nu\gamma,rs} \bar{D}_{\alpha\nu}^R (\omega,m,r) \bar{\Pi}^<_{\nu\gamma} \left(\omega,r,s\right) \bar{D}_{\gamma\beta}^A \left(\omega,s,n\right).
\end{split}
\end{equation}
Other important objects transform in a similar manner. The lead self-energies, equation (\ref{eq:lead_self_energies}), making use of equations (\ref{eq:jacobi_anger}) and (\ref{eq:floquet_convolution}), transform to
\begin{equation}
	\bar{\Sigma}_{\alpha,ij} \left(\omega,m,n\right) = \sum_{pq,lk} \bar{S}_{\alpha}\left(m,p\right) {\bar{\Sigma}'}_{\alpha,lk} \left(\omega,p,q\right)  {\bar{S}}_{\alpha} \left(n,q\right),
\end{equation}
where $\bar{S}_{\alpha}\left(m,n\right) = J_{m-n} \left(\Delta_\alpha / \Omega_\alpha \right)$. The Fourier coefficients of the current and occupation can be taken from the Floquet matrices derived from equations (\ref{eq:current}) and (\ref{eq:electronic_occupation}):
\begin{equation}
	\begin{split}
		I_\alpha(n-m) = \bar{I}_{\alpha}\left(m,n\right) \\
		=  2 \int^{\infty}_{-\infty} \frac{d\omega}{2\pi}  \sum_{r,ij} \left[ \bar{G}_{ij}^R (\omega,m,r) \bar{\Sigma}^<_{ji} \left(\omega,r,n\right)  \right.
\\
\left. + \bar{G}_{ij}^< (\omega,m,r) \bar{\Sigma}^R_{ji} \left(\omega,r,n\right) \right]
	\end{split}
\end{equation}
and
\begin{equation}\label{eq:electronic_occupation_floquet}
	\begin{split}
		n_{j}^{el} \left(n-m\right) = \bar{n}_{j}^{el} \left(m,n\right) = -i \int^{\infty}_{-\infty} \frac{d\omega}{2\pi}  \bar{G}_{jj}^< (\omega,m,n),
	\end{split}
\end{equation}
The time-resolved observables can then be found with equation (\ref{eq:periodicity_1}), where $t\rightarrow t'$, and by taking only the real part, as per equations (\ref{eq:electronic_occupation}) and (\ref{eq:current}).

\begin{figure*}[t!]
	\centering
	\hspace*{-7cm} 
	\begin{subfigure}[]{1in}
		\centering
		\includegraphics[width=3.6\textwidth]{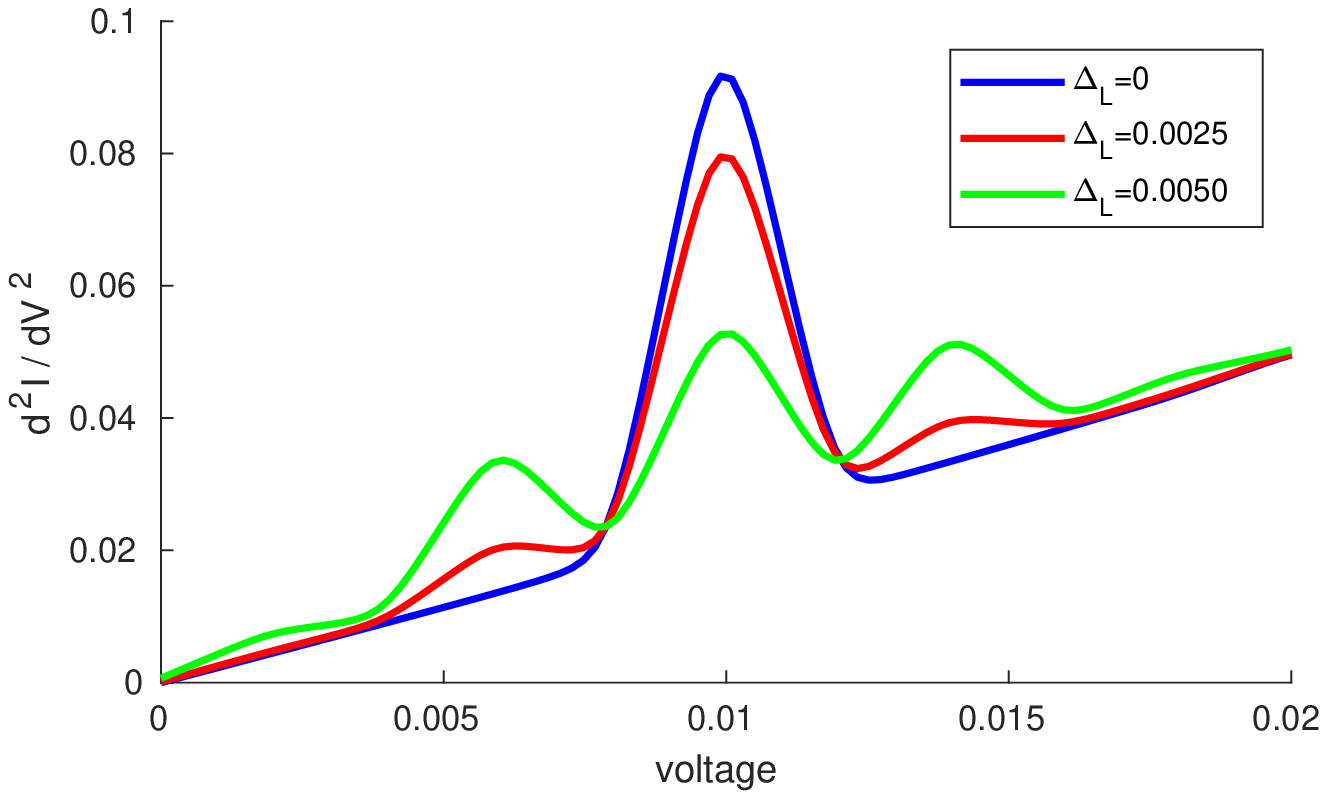} 
		\caption{\label{fig:plot1a}}
	\end{subfigure}   
	\qquad\qquad\qquad\qquad\qquad\qquad\qquad\qquad\qquad\qquad\quad
	\begin{subfigure}[]{1in}
		\centering
		\includegraphics[width=3.6\textwidth]{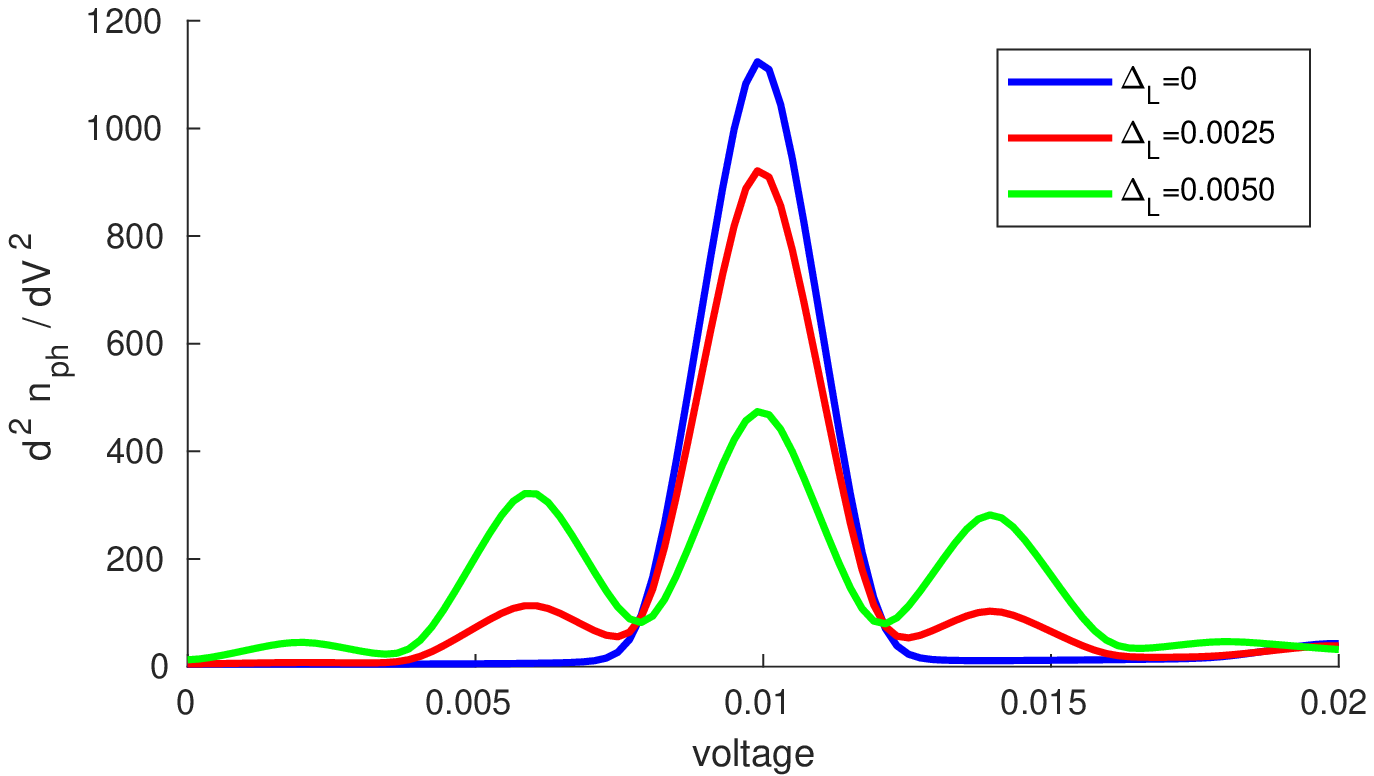} 
		\caption{\label{fig:plot1b}}
	\end{subfigure}	
	\caption{The changes in $d^2I/dV^2$ and $d^2n_{ph}/dV^2$ as voltage increases, given different driving energies $\Delta_L$. The other parameters are $\Gamma_L=\Gamma_R=0.015$, $\epsilon_c=0.1$, $\eta_c = 3\times 10^{-5}$, $\Omega=0.004$, $\omega_c=0.01$, $\lambda_c=0.015$ and $T=1.5\times10^{-4}$. The bounds of the integrand were taken at $-0.3$ and $0.3$. Fourier coefficients ranging from -8 to 8 were used in the calculation. The uniform grid spacing was $2\times10^{-5}$. The convergence was below $10^{-6}$ for both the electronic and phonon occupation.}
	\label{fig:plot1}
\end{figure*}

Unfortunately, the interaction self-energies cannot be brought to such an amenable form, and must be calculated from the Fourier coefficients. The interaction self-energies follow the forms
\begin{equation}
C_{XC} (t,t')  = A\left(t,t'\right) B\left(t,t'\right),
\end{equation}
\begin{equation}
C_{POL} (t,t')  = A\left(t,t'\right) B\left(t',t\right),
\end{equation}
\begin{equation}
C_{HAR} (t,t') = \delta \left(t-t'\right) \int^{\infty}_{-\infty} dt_1 A (t,t_1) B (t_1,t_1).
\end{equation}
Applying the Floquet tranformation to the above gives
\begin{equation}
\begin{split}
C_{XC} \left(\omega,n\right) 
\\
= \sum_{m = -\infty}^{\infty} \int \frac{d\omega'}{2\pi} A(\omega',m) B(\omega-\omega',n-m),
\end{split}
\end{equation}
\begin{equation}
\begin{split}
C_{POL} \left(\omega,n\right) 
\\
= \sum_{m = -\infty}^{\infty} \int \frac{d\omega'}{2\pi} A(\omega',m) B(\omega'-\omega,n-m)
\end{split}
\end{equation}
and
\begin{equation}
\begin{split}
C_{HAR} (\omega,n) 
\\
= \sum^{\infty}_{m=-\infty} A \left(-\frac{\Omega}{2} \left(n+m\right),n-m\right) B' \left(n\right).
\end{split}
\end{equation}
The $C_{HAR}\left(\omega,n\right)$ simplifies when $A(t,t')=A(t-t')$, giving us 
\begin{equation}
\begin{split}
C_{HAR}\left(\omega,n\right)
\\
= \sum^{\infty}_{m=-\infty} \delta_{m,n} A(-\frac{\Omega}{2}\left(m+n\right),0) B(n) 
\\
= A(-\Omega n,0) B(n).
\end{split}
\end{equation}
The above allow us to cast the interaction self-energies in terms of their Fourier coefficients:
\begin{equation}
\begin{split}
\Sigma^R_{har,ij} \left(\omega,r\right) =
\\ -i \sum_{\beta}\lambda^\beta_{ij} d^R_{\beta}(\omega=-\Omega r) \int \frac{d\omega}{2\pi} \sum_{kw} \lambda_{kw}^{\beta} G_{wk}^< (\omega,r)
\end{split}
\end{equation}
\begin{equation}
\begin{split}
\Sigma^R_{XC,ij} \left(\omega,r\right) =
\int \frac{d\omega'}{2\pi} \sum^{\infty}_{n=-\infty} \sum_{\mu\nu, ml}
\\ iD^<_{\mu\nu}\left(\omega',n\right)\lambda^{\mu}_{im} G^R_{ml}\left(\omega-\omega',r-n\right) \lambda^{\nu}_{lj} 
\\
+ iD^R_{\mu\nu}\left(\omega',n\right)\lambda^{\mu}_{im} G^<_{ml}\left(\omega-\omega',r-n\right) \lambda^{\nu}_{lj} 
\\
+iD^R_{\mu\nu}\left(\omega',n\right)\lambda^{\mu}_{im} G^R_{ml}\left(\omega-\omega',r-n\right) \lambda^{\nu}_{lj} 
\end{split}
\end{equation}
\begin{equation}
\begin{split}
\Sigma^<_{XC,ij} \left(\omega,r\right) = \int \frac{d\omega'}{2\pi} \sum^{\infty}_{n=-\infty} \sum_{\mu\nu, ml} 
\\
iD^<_{\mu\nu}\left(\omega',n\right)\lambda^{\mu}_{im} G^<_{ml}\left(\omega-\omega',r-n\right) \lambda^{\nu}_{lj} 
\end{split}
\end{equation}
\begin{equation}
\begin{split}
\Sigma^>_{XC,ij} \left(\omega,r\right) = \int\frac{d\omega'}{2\pi} \sum^{\infty}_{n=-\infty} \sum_{\mu\nu, ml}
\\ iD^>_{\mu\nu}\left(\omega',n\right)\lambda^{\mu}_{im} G^>_{ml}\left(\omega-\omega',r-n\right) \lambda^{\nu}_{lj} 
\end{split}
\end{equation}
\begin{equation}
\begin{split}
\Pi^R_{\alpha\beta} (\omega,r) = \int \frac{d\omega'}{2\pi}\sum^{\infty}_{n=-\infty} \sum_{mlkp}
\\ -i \lambda^{\alpha}_{ml} G^<_{lk}\left(\omega',n\right) \lambda^{\beta}_{kp} G^A_{pm}\left(\omega'-\omega,r-n\right)
\\
-i\lambda^{\alpha}_{ml} G^R_{lk}\left(\omega',n\right) \lambda^{\beta}_{kp} G^<_{pm}\left(\omega'-\omega,r-n\right)
\end{split}
\end{equation}
\begin{equation}
\begin{split}
\Pi_{\alpha\beta}^< (\omega,r) = \int \frac{d\omega'}{2\pi}\sum^{\infty}_{n=-\infty} \sum_{mlkp}
\\ -i \lambda^{\alpha}_{ml} G^<_{lk}\left(\omega',n\right) \lambda^{\beta}_{kp} G^>_{pm}\left(\omega'-\omega,r-n\right)
\end{split}
\end{equation}
\begin{equation}
\begin{split}
\Pi_{\alpha\beta}^> (\omega,r) = \int \frac{d\omega'}{2\pi}\sum^{\infty}_{n=-\infty} \sum_{mlkp} \\ -i \lambda^{\alpha}_{ml} G^>_{lk}\left(\omega',n\right) \lambda^{\beta}_{kp} G^<_{pm}\left(\omega'-\omega,r-n\right).
\end{split}
\end{equation}
Solving equation (\ref{eq:position_eom}), we can cast the average phonon positions, and the subsequent average phonon momenta, in terms of Fourier coefficients
\begin{equation}\label{eq:position}
\begin{split}
\left\langle Q_{\alpha} \right\rangle (r) 
\\
=-i   d^R_\alpha \left(\omega=-\Omega r\right) \int \frac{d\omega}{2\pi} \sum_{ml}\lambda_{ml}^\alpha G_{lm}^<\left(\omega,r\right),
\end{split}
\end{equation}
\begin{equation}
\left\langle P_{\alpha} \right\rangle (r) = \frac{ir\Omega}{\omega_0}  \left\langle Q_{\alpha} \right\rangle (r).
\end{equation}
We can complete a similar process for the phonon momentum Green's functions, allowing us to calculate variance of the momentum operators:
\begin{equation}
\begin{split}
\left\langle \left(\Delta P_{\alpha} \right)^2 \right\rangle (r) = iD_{\alpha\alpha}^{PP,<} (r)
\\
= \frac{i}{{\omega_0}^2} \int \frac{d\omega}{2\pi} \left(\omega^2 - \left(\frac{r\Omega}{2}\right)^2\right) D_{\alpha\alpha}^< (\omega,r).
\end{split}
\end{equation}

\subsection{Implementation}

Solving for the Green's functions, the dimensions of the Floquet matrices, and the corresponding Fourier series, need to be truncated. The addition of more Fourier coefficients leads more accurate results, converging on the exact result. Calculating the integrand of the Fourier coefficients was completed with equidistant grid of points. Completing this procedure for the noninteracting case, the Floquet matrices were unravelled to Fourier coefficients using equation (\ref{eqn:floquet_matrix}). The terms where $n+m=0,-1$ of $\bar{A}\left(\omega,m,n\right)$, where taken for calculating the Fourier coefficients. These were then used to calculate the interaction self-energies before being reassembled into the Floquet matrices. The process was then completed iteratively, with convergence given by the Fourier coefficients of the phonon and electron occupations:
\begin{equation}
\begin{split}
	\frac{\sum_{m} \left| n_m^{k+1} - n_m^{k} \right|}{\sum_{m} \left|n^k_m\right|} \leq \delta,
\end{split}
\end {equation}
where $n^k_m$ is the $k$th iteration of the $m$th Fourier coefficent of the occupation in question, with $\delta$ as the convergence.

\begin{figure*}[t!]
	\centering
	\hspace*{-7cm} 
	\begin{subfigure}[]{1in}
		\centering
		\includegraphics[width=3.6\textwidth]{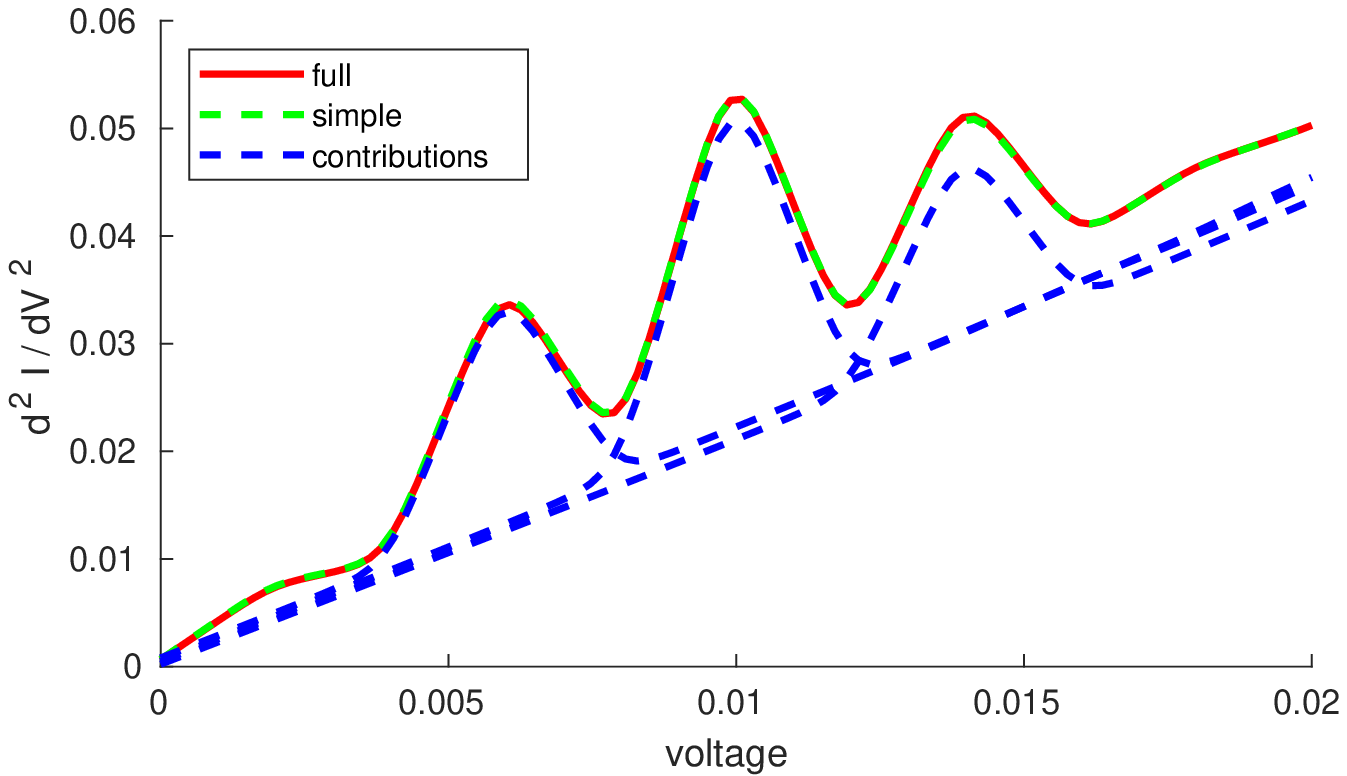} 
		\caption{\label{fig:plot2a}}
	\end{subfigure}   
	\qquad\qquad\qquad\qquad\qquad\qquad\qquad\qquad\qquad\qquad\quad
	\begin{subfigure}[]{1in}
		\centering
		\includegraphics[width=3.6\textwidth]{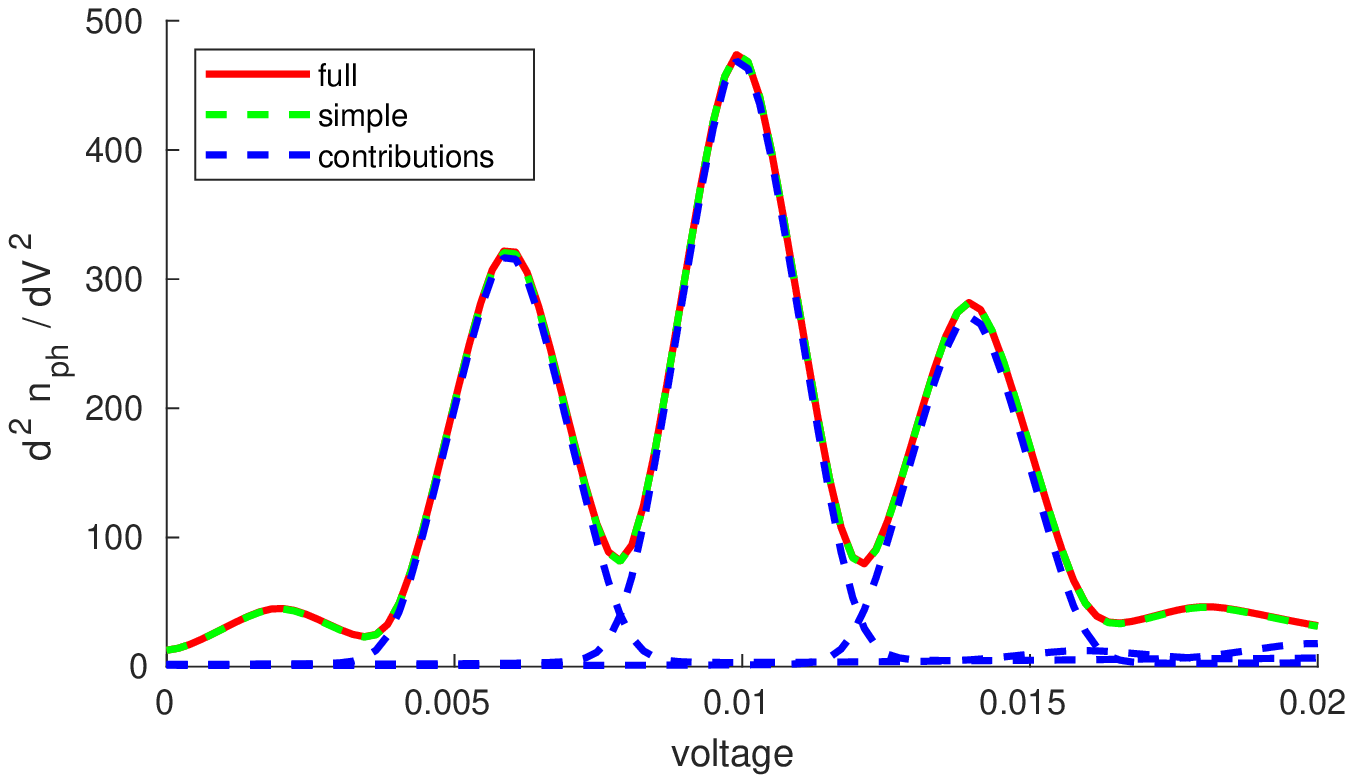} 
		\caption{\label{fig:plot2b}}
	\end{subfigure}	
	\caption{The changes in $d^2I/dV^2$ and $d^2n_{ph}/dV^2$ as voltage increases. Here, the full method is plotted alongside the simplistic method and contributions from $n=-1,0,1$ of equation (\ref{eq:simplistic_model}). The other parameters are $\Gamma_L=\Gamma_R=0.015$, $\epsilon_c=0.1$, $\eta_c = 3\times10^{-5}$, $\Omega=0.004$, $\omega_c=0.01$, $\lambda_c=0.015$, $\Delta_L=0.005$ and $T=1.5\times10^{-4}$. The bounds of the integrands were taken at $-0.3$ and $0.3$. Fourier coefficients ranging from -8 to 8 were used in the calculation. The uniform grid spacing was $2\times10^{-5}$. The convergence was below $10^{-6}$ for both the electronic and phonon occupations.}
	\label{fig:plot2}
\end{figure*}

\section{Results}

For simplicity, we focus on a single electronic level coupled to a single phonon mode, with driving within the left lead:
\begin{equation}
	\begin{split}
	H\left(t\right) = \epsilon_c d^\dagger d + \sum_{i k \alpha=L,R } t_{k\alpha i} \; c^{\dagger}_{k \alpha} d +  t^*_{k\alpha i} \; d^\dagger c_{k \alpha} 
	\\
	+\sum_{k}  \left(\epsilon_{kL} + \Delta_L \cos (\Omega t) \right) c^{\dagger}_{kL} c_{kL} + \epsilon_{kR} c^{\dagger}_{kR} c_{kR}
	\\
	+ \; \omega_c a^\dagger a + \lambda_c Q d^\dagger d.
	\end{split}
\end{equation}
The parameters for the model consist of $\epsilon_c$, the energy of the central level, $\omega_c$, the vibrational frequency of the phonon mode, $\Omega$, the driving frequency of the left lead, $\Delta_L$ the magnitude of the driving in the left lead, $\Gamma_{L/R}$, the couplings to the leads, $\lambda_c$, the coupling strength between the central level and phonon, $\eta_c$, the coupling of the phonon to its bath and the temperature of both leads, $T$. Within the calculations, the driving frequency of the left lead is used as the frequency for the system's periodicity.

For the time-averaged picture of this model, we have two important limiting cases in the static, interacting case and the noninteracting case. The static case is well understood and extensively studied\cite{Galperin2004a,Galperin2004b,Galperin2007,cuevas2010molecular}. In this context, the phonon mode causes elastic corrections and facilitates new, inelastic channels of transport through the junction, by means of absorption or emission of phonons. The latter only occurs when the voltage window widens to accommodate electrons that enter the junction before absorbing (emitting) a phonon of energy. The addition of extra channels through junction can be seen in subtle changes to the current, captured as peaks in the derivative of the differential conductance with respect to voltage.

The noninteracting case has also been extensively studied, and can be explained with the notion of photon-assisted transport \cite{Jauho1994,Haug2008,Platero2004}. The periodic driving of the system (environment) results in contributions to the time-averaged observables from the equivalent static cases, with the driven energies shifted by integer multiples of the driving frequency of the time-dependence. This can be interpreted as a proportion of the electrons emitting or absorbing quanta of the energy, hence the name photon-assisted transport. For the driving used in this paper, see equation (\ref{eq:phase_definition}), limited to the left lead, we have
\begin{equation}\label{eq:simplistic_model}
\begin{split}
\frac{1}{P} \int^P_{0} I_\alpha(t) dt = 
\\
\sum^{n=\infty}_{n=-\infty} \left( J_n \left(\frac{\Delta_\alpha}{\Omega_\alpha}\right)  \right)^2 \; I^{DC}_{\alpha} \left(\mu_L+n\Omega,\mu_R\right),
\end{split}
\end{equation} 
Where, $I_\alpha(t)$ is the AC-driven current through lead $\alpha$ and $I^{DC}_a$ is the static case, where no driving is present.

Within certain parameter regimes, combining the reasoning from both limiting cases explains the features observed in the full model. In figure (\ref{fig:plot1a}), we see the primary peak within $d I^2 / d V^2$, indicative of inelastic collisions, gain additional satellite peaks due to absorption (emission) of quanta of energy by the electrons, prior to entering the junction, per photon-assisted tunnelling. A similar effect is observed within figure (\ref{fig:plot1b}), with $d^2 n_{ph}/dV^2$ gaining photon-assisted sidepeaks, suggesting that the photon-assisted side-peaks of left lead contribute to the occupation of the phonon independently of each other.

\begin{figure*}[pt!]
	\centering
	\hspace*{-7cm} 
	\begin{subfigure}[]{1in}
		\centering
		\includegraphics[width=3.6\textwidth]{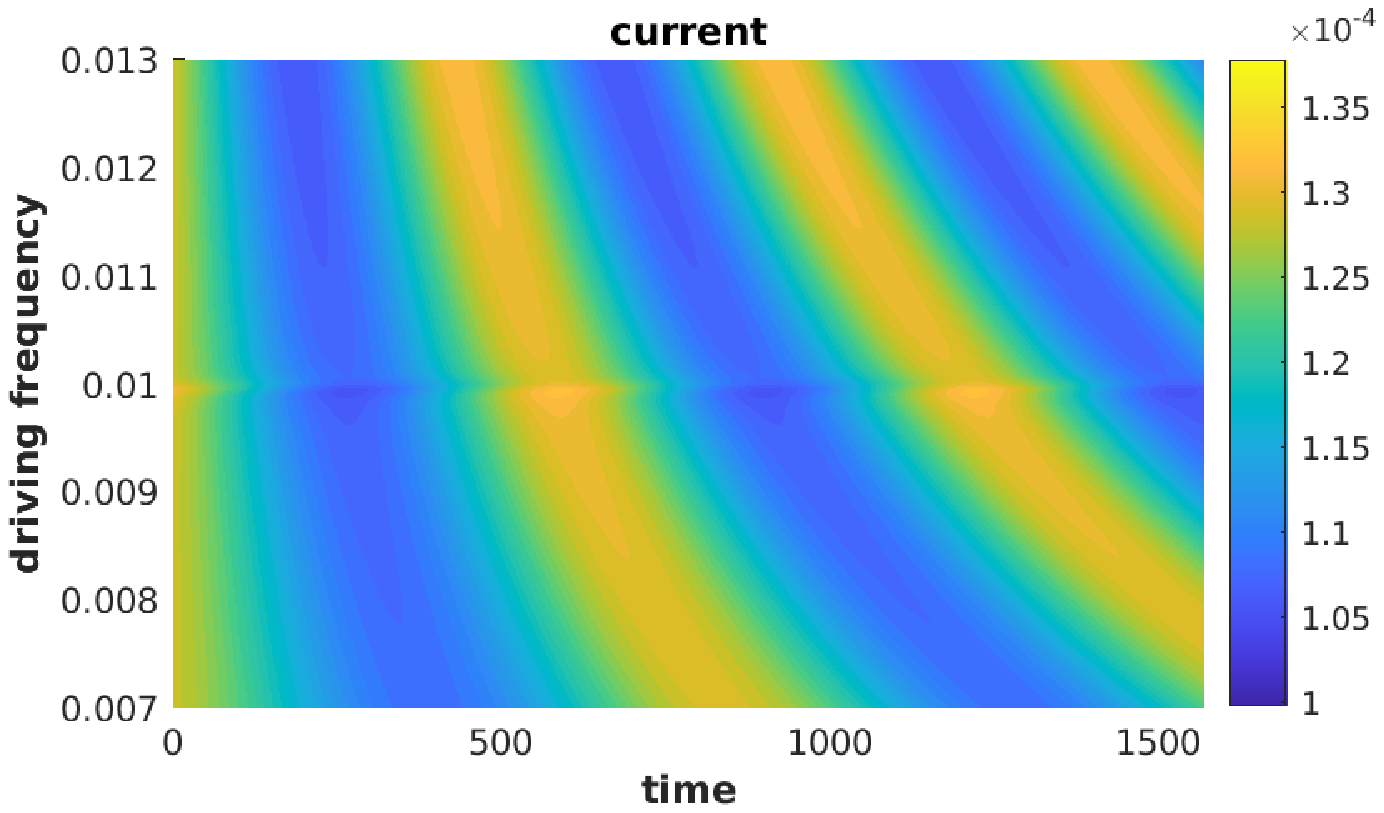} 
		\caption{}
	\end{subfigure}   
	\qquad\qquad\qquad\qquad\qquad\qquad\qquad\qquad\qquad\qquad\quad
	\begin{subfigure}[]{1in}
		\centering
		\includegraphics[width=3.6\textwidth]{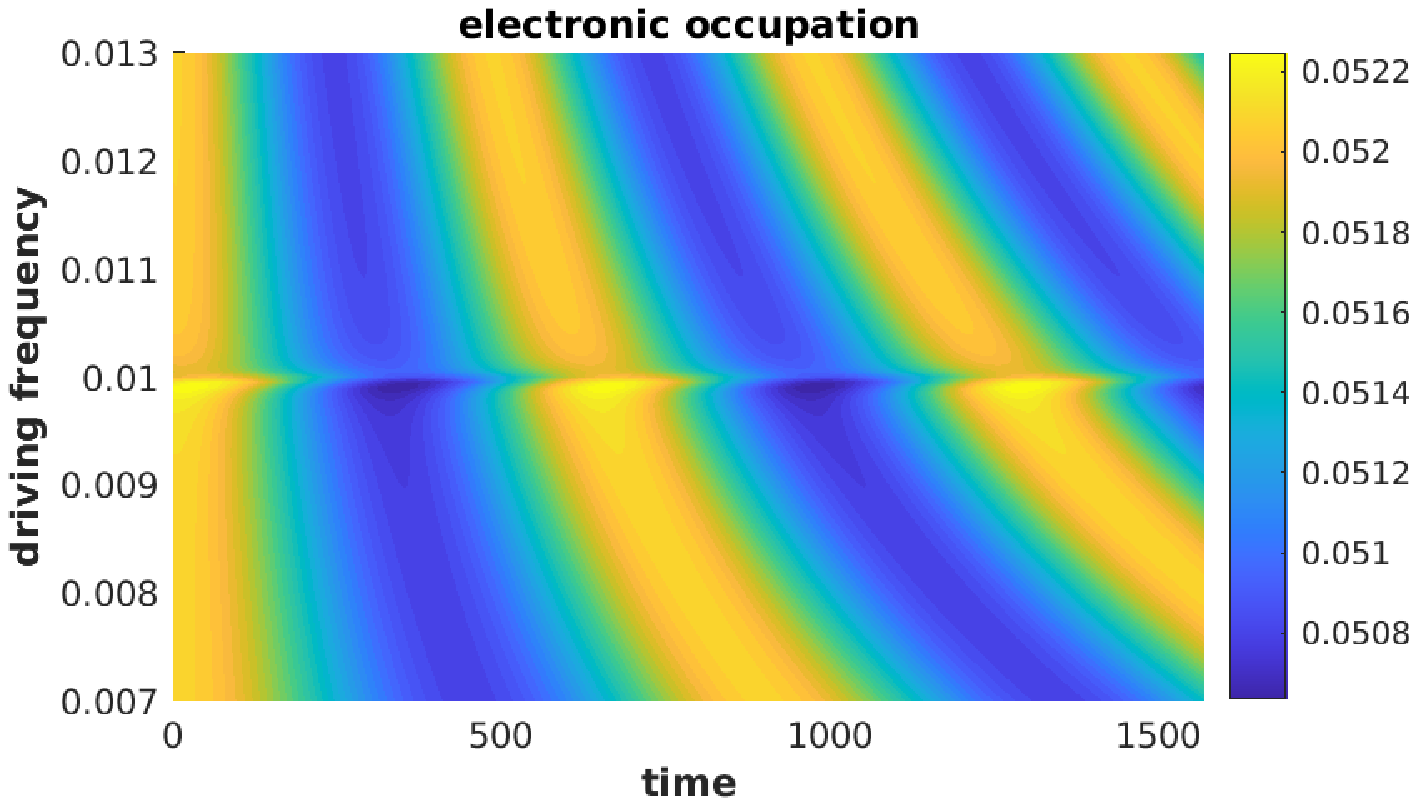} 
		\caption{}
	\end{subfigure}
	
	\hspace*{-7cm} 
	\begin{subfigure}[]{1in}
		\centering
		\includegraphics[width=3.6\textwidth]{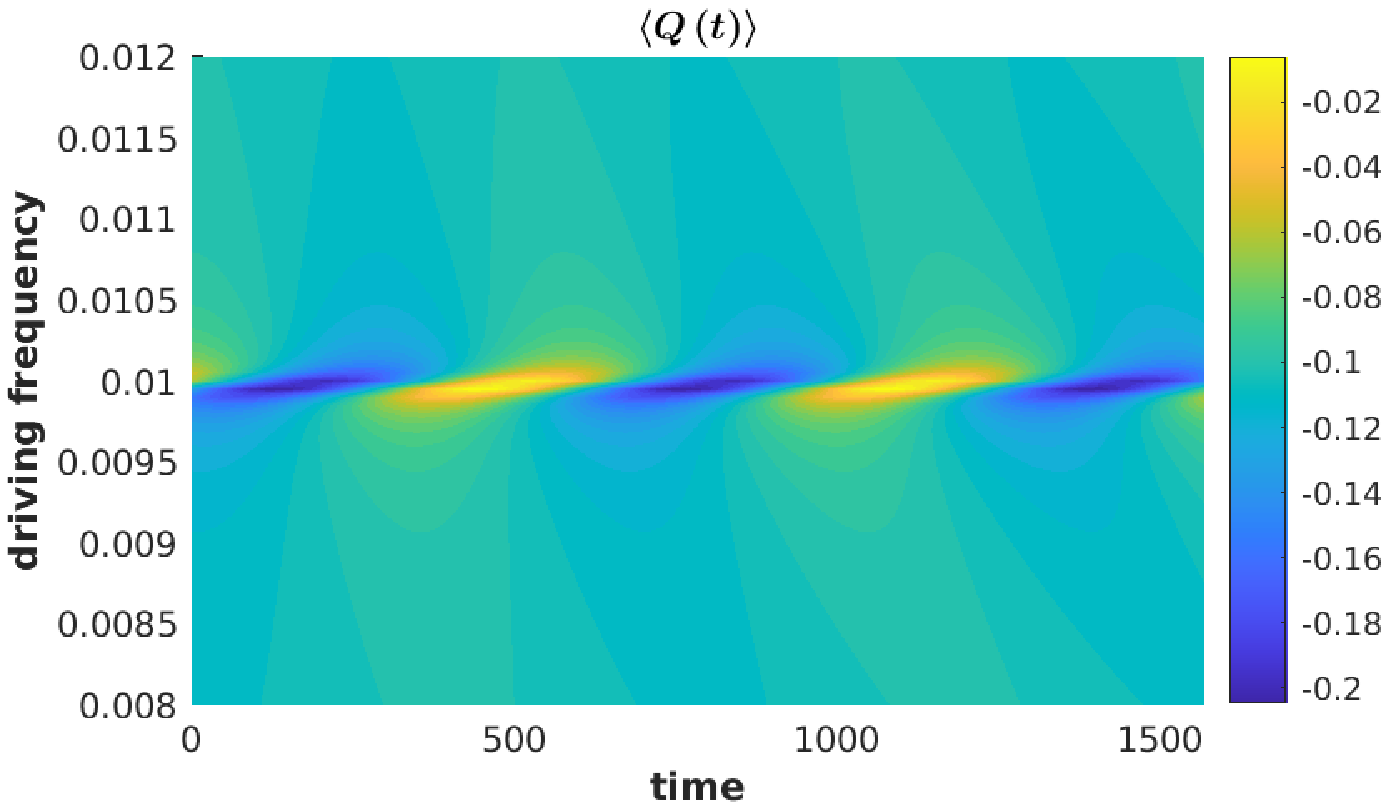} 
		\caption{}
		\label{fig:resonance_position}
	\end{subfigure}   
	\qquad\qquad\qquad\qquad\qquad\qquad\qquad\qquad\qquad\qquad\quad
	\begin{subfigure}[]{1in}
		\centering
		\includegraphics[width=3.6\textwidth]{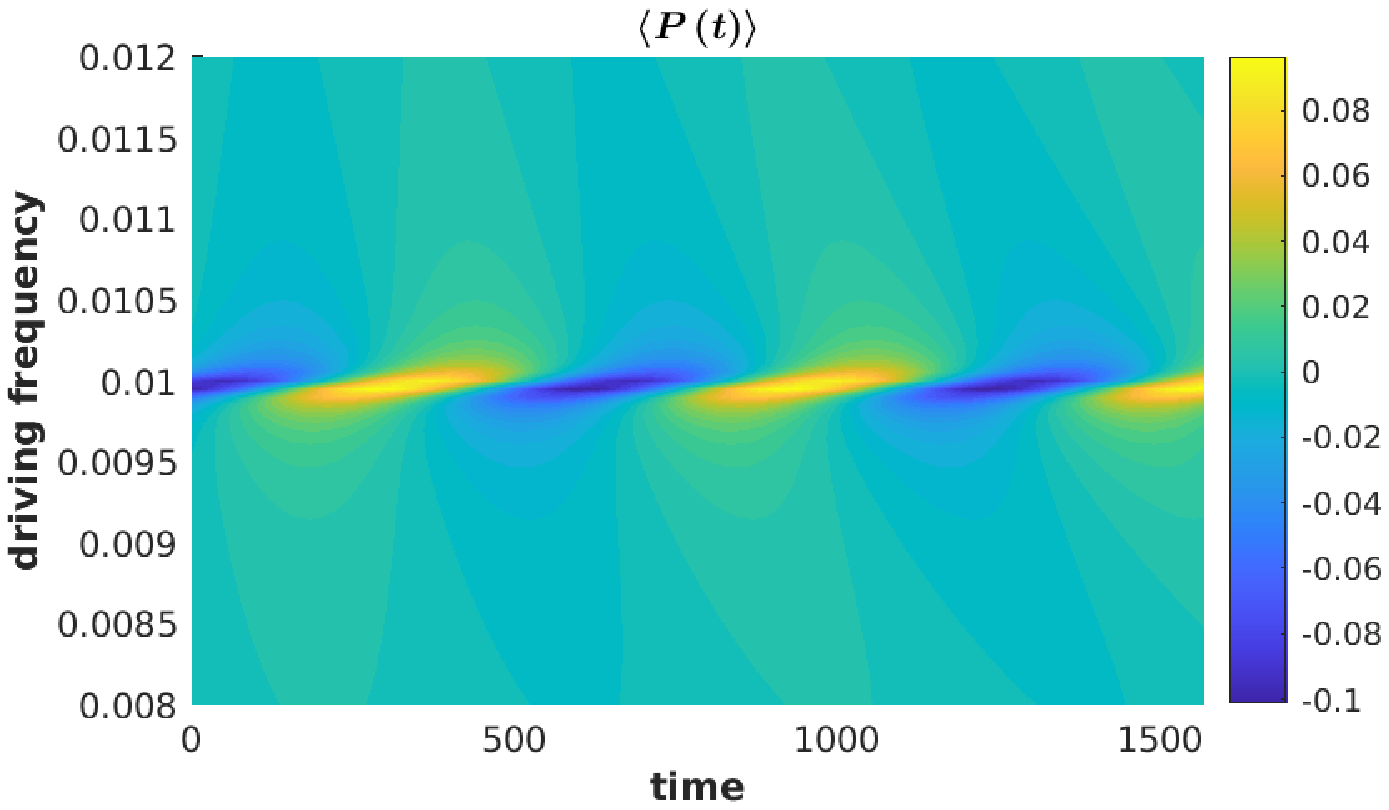} 
		\caption{}
	\end{subfigure}
	
	\hspace*{-7cm} 
	\begin{subfigure}[]{1in}
		\centering
		\includegraphics[width=3.6\textwidth]{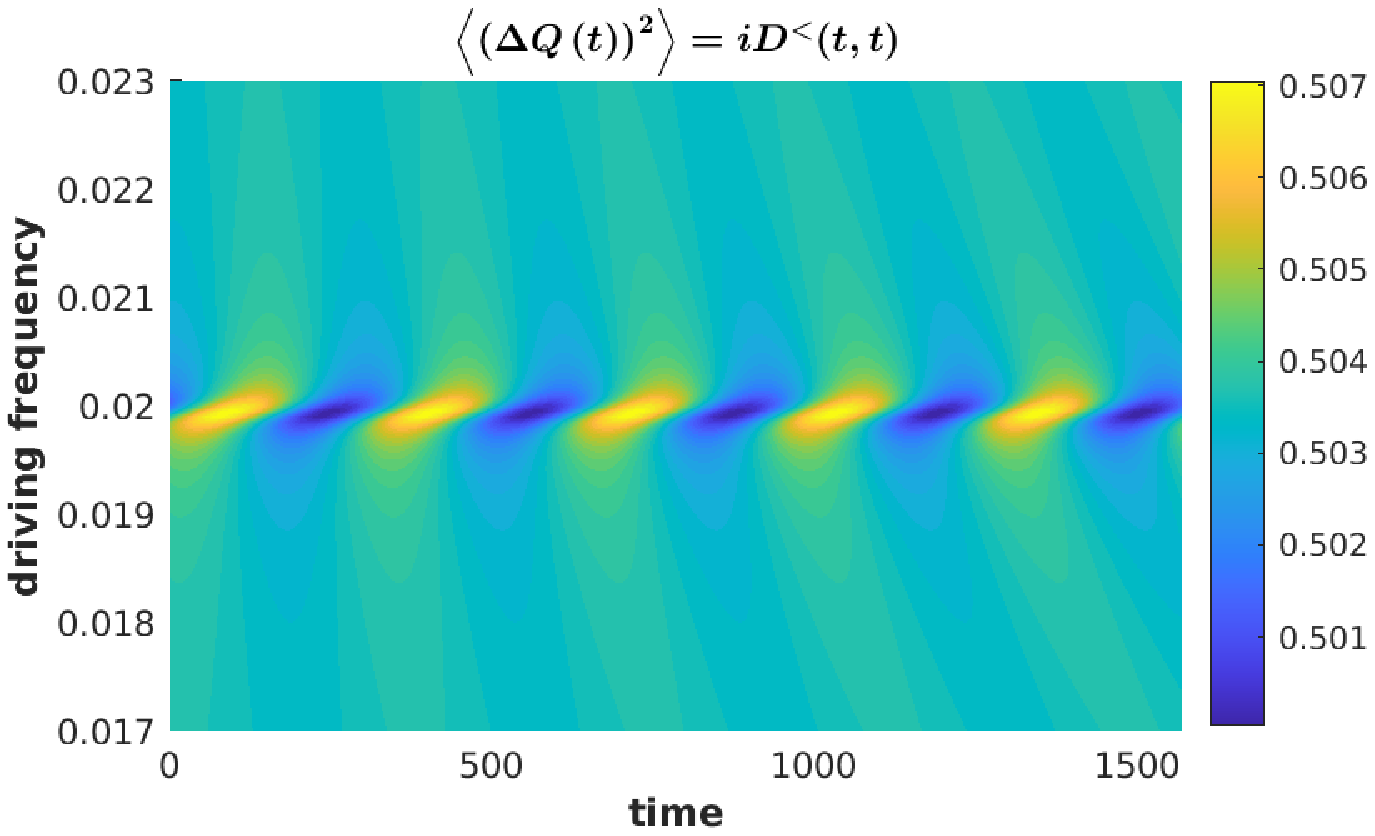} 
		\caption{}
		\label{fig:resonance_e}
	\end{subfigure}   
	\qquad\qquad\qquad\qquad\qquad\qquad\qquad\qquad\qquad\qquad\quad
	\begin{subfigure}[]{1in}
		\centering
		\includegraphics[width=3.6\textwidth]{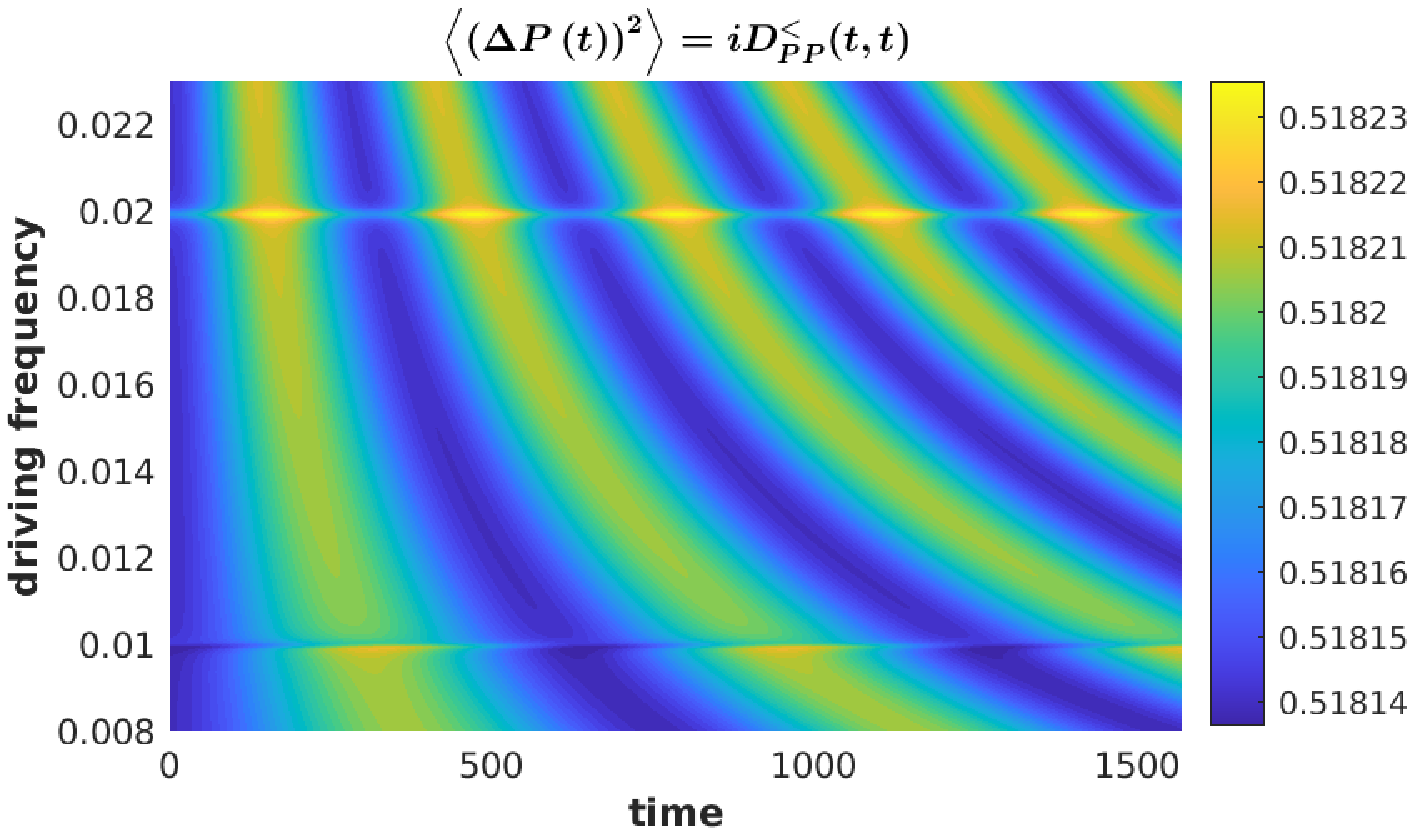} 
		\caption{}
		\label{fig:resonance_f}
	\end{subfigure}
	
	\hspace*{-7cm} 
	\begin{subfigure}[]{1in}
		\centering
		\includegraphics[width=3.6\textwidth]{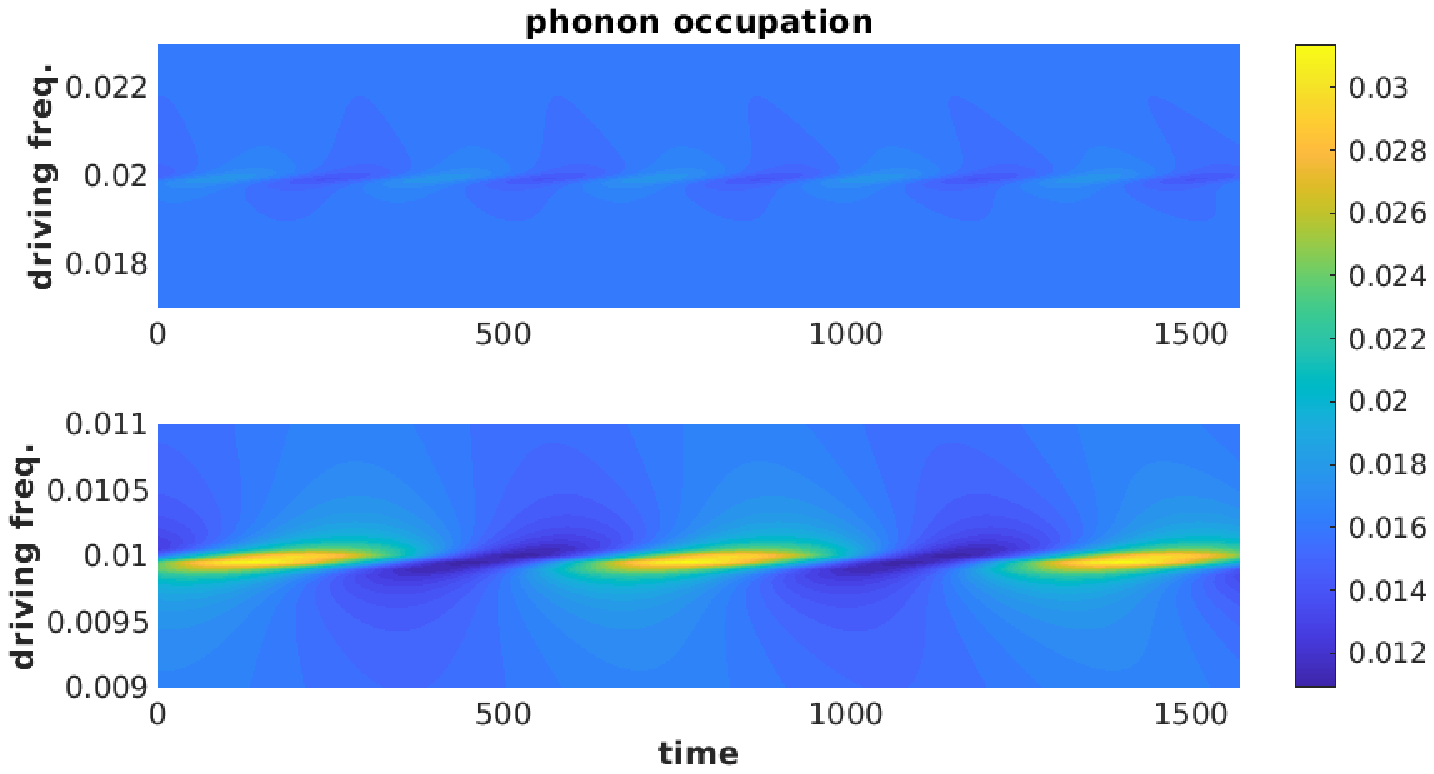} 
		\caption{}
		\label{fig:resonance_g}
	\end{subfigure}   
	\qquad\qquad\qquad\qquad\qquad\qquad\qquad\qquad\qquad\qquad\quad
	\begin{subfigure}[]{1in}
		\centering
		\includegraphics[width=3.6\textwidth]{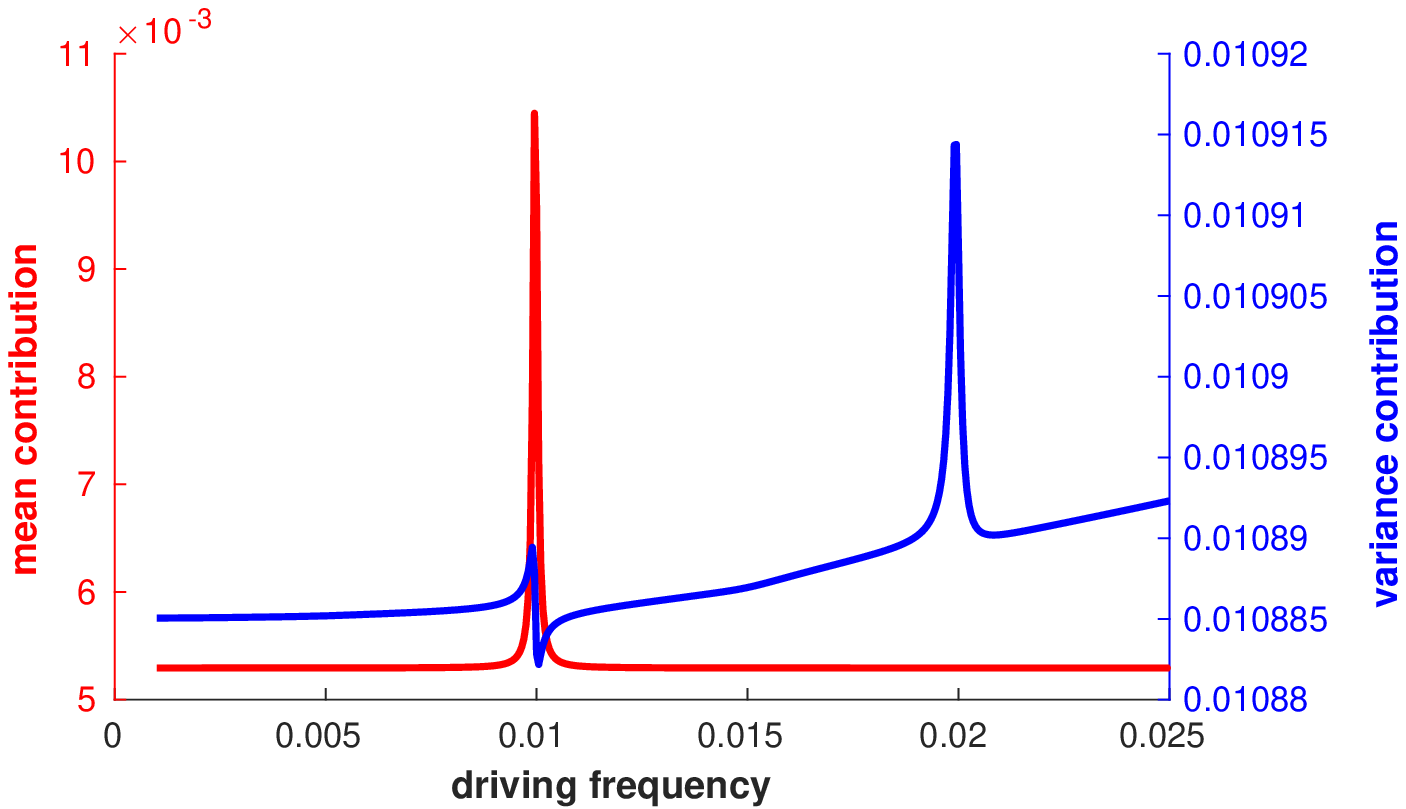} 
		\caption{}
		\label{fig:resonance_h}
	\end{subfigure}
	
	\caption{Figures (a) to (g) plot objects of interest over time, whilst figure (h) plots the time-averaged contributions to the phonon occupation, as given by equation (\ref{eq:phonon_occupation}). The physical parameters are $\Gamma_L=\Gamma_R=0.015$, $\epsilon_c = 0.1$, $T=1.5\times10^{-4}$, $\omega_c=0.01$, $\eta_c=6\times10^{-5}$, $\Delta_L=0.0015$ and $\lambda_c=0.01$. Fourier coefficients ranging from -14 to 14 were used in the calculation, with an integrand discretization of $2.5\times10^{-5}$ with bounds of -1 and 1. The convergence was below $10^{-4}$ for both the electronic and phonon occupation.}
	\label{fig:resonance}
\end{figure*}

The above insights suggest a simplistic model where equation (\ref{eq:simplistic_model}) is augmented, with the static components being calculated with the addition of electron-phonon interactions. This method was often found to successfully predict the inelastic features of the full model. See figure (\ref{fig:plot2}) for an example, where the contributions given by equation (\ref{eq:simplistic_model}) are plotted alongside the full and simplistic methods. The convergence for the simplistic model was calculated by using the average occupation as if it were static.

\begin{figure*}[t!]
	\centering
	\hspace*{-4cm} 
	\begin{subfigure}[]{0.66in}
		\includegraphics[width=3.6\textwidth]{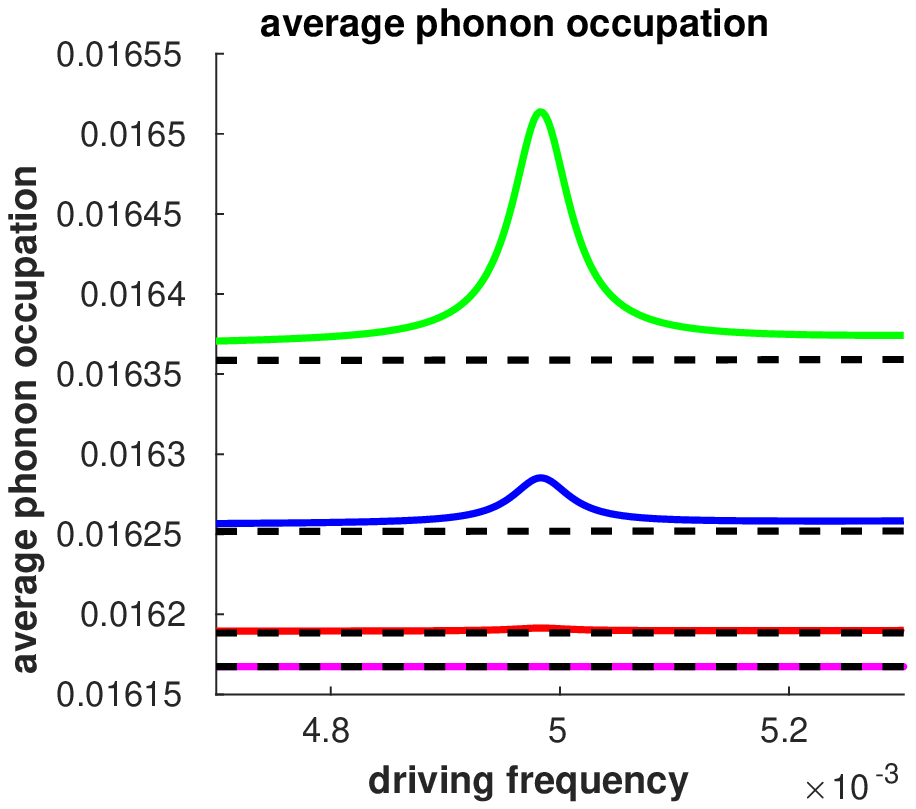} 
		\caption{}
		\label{fig:plot4a}
	\end{subfigure}   
	\qquad\qquad\qquad\qquad\qquad\qquad
	\begin{subfigure}[]{0.66in}
		\includegraphics[width=3.6\textwidth]{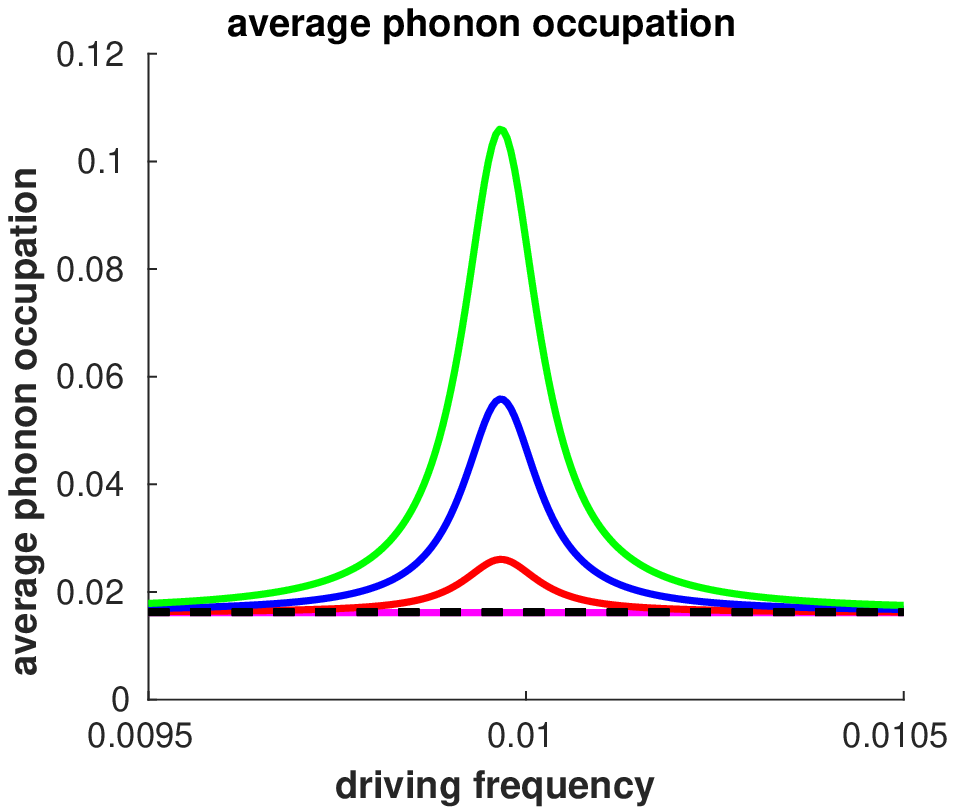} 
		\caption{}
		\label{fig:plot4b}
	\end{subfigure}
	\qquad\qquad\qquad\qquad\qquad\qquad
	\begin{subfigure}[]{0.66in}
		\includegraphics[width=3.6\textwidth]{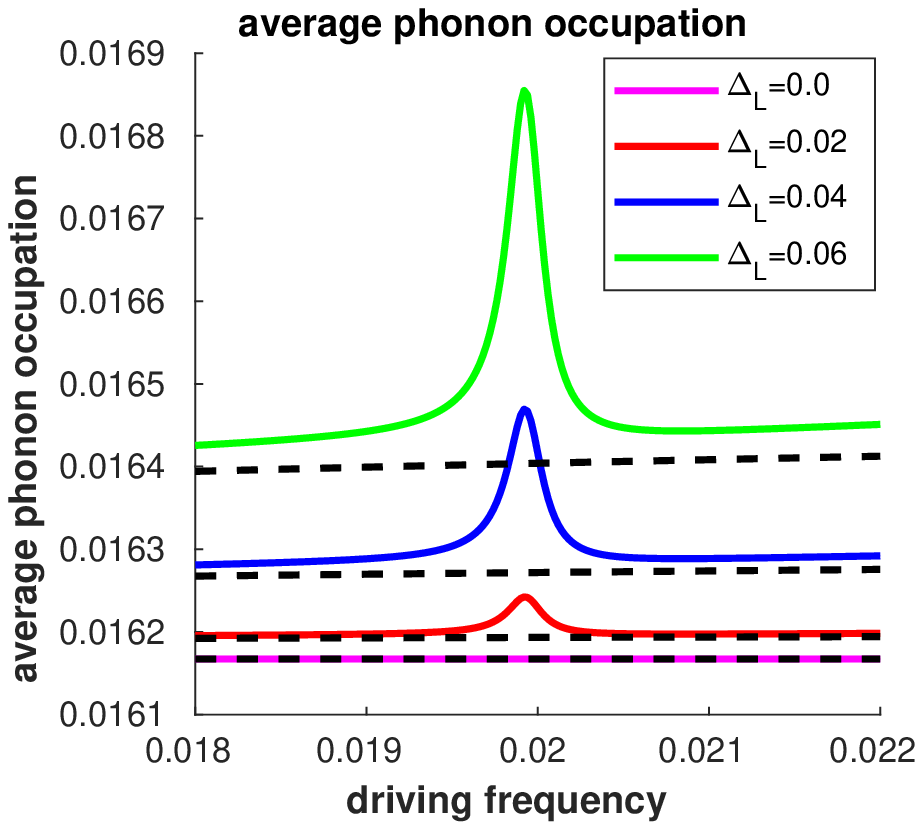} 
		\caption{}
		\label{fig:plot4c}
	\end{subfigure}
	\caption{The time-averaged phonon occupation as driving frequency increases. Here, the driving energies of the left lead, $\Delta_L$, have been varied. Furthermore, the simplistic method has been plotted in dashed black. The other parameters are $\Gamma_L=\Gamma_R=0.015$, $\epsilon_c=0.1$, $\eta_c = 6\times10^5$, $\omega_c=0.01$, $\lambda_c=0.01$ and $T=1.5\times10^{-4}$. The bounds of the integrands was taken at -1 and 1. Fourier coefficients ranging from -8 to 8 were used in the calculation. The uniform grid spacing was $5\times10^{-6}$ for plots (a) and (b) whilst plot (c) was calculated with $1\times10^{-5}$. The convergence was below $10^{-4}$ for both the electronic and phonon occupation.}
	\label{fig:plot4}
\end{figure*}

The simplistic model can be motivated for situations where the timescales for interaction between the electronic and phononic components, $t_\lambda \sim 1/\lambda_c$, is far longer than the traversal time for the electrons within the junction\cite{cuevas2010molecular}:
\begin{equation}
 \frac{1}{\lambda_c} \gg \frac{1}{\sqrt{\Gamma^2 + {\Delta E}^2}},
\end{equation} 
where $\Gamma = \Gamma_L + \Gamma_R$ and $\Delta E$, the injection energy, is the distance of the energy level from resonance, usually taken as the difference between the energy level and closest chemical potential.
In the regime specified by the above assumption, the phonon mode will see the time-dependence of the electronic component averaged over the long interaction time, hence the ability of the simplistic model to capture the dynamics. This insight is similar to that used to investigate the effects of AC driving with master equations \cite{Peskin2017,Peskin2020}, where weak coupling between central region and leads results in the central region seeing an average picture of the leads' dynamics.

Within the model in question, it was found that resonance driving at $\Omega\approx\omega_c$ resulted in significant variations for several parameter ranges (see figures (\ref{fig:resonance}) and (\ref{fig:plot4})). This is mostly due to the sensitivity of average position, to resonant driving, which in turn influences the other observables. This can be seen in figure (\ref{fig:resonance_position}). This sensitivity to resonance comes from equation (\ref{eq:position}), for which the single level case simplifies to
\begin{equation}\label{eq:single_level_position}
\begin{split}
\left\langle Q \right\rangle (m) 
\\
= \lambda_c \; d^R \left(-\Omega m\right) n_{el}(m).
\end{split}
\end{equation}

Focusing on the bare, retarded phonon Green's function, we can separate out the real and imaginary parts:
\begin{equation}\label{eq:single_level_dr}
\begin{split}
d^R(\omega)  =
\\ 
\frac{i}{2}\left(\frac{-\eta_c}{\left(\omega-\omega_c\right)^2 + \eta_c^2} + \frac{\eta_c}{\left(\omega+\omega_c\right)^2 + \eta_c^2}\right)
\\
+\frac{1}{2}\left(\frac{\omega-\omega_c}{\left(\omega-\omega_c\right)^2 + \eta_c^2} + \frac{\omega_c+\omega}{\left(\omega_c+\omega\right)^2 + \eta_c^2}\right).
\end{split}
\end{equation}
The real and imaginary components of above are maximised around $\pm\Omega n \approx \omega_c$, especially when the $\eta_c \ll \omega_c$. This results in the average phonon position being sensitive to periodic variation in the electronic occupation, resulting in the primary resonance peak, seen in figures (\ref{fig:resonance_g}), (\ref{fig:resonance_h}) and (\ref{fig:plot4b}), around $\Omega\approx\omega_c$. Additionally, smaller subharmonic resonances can also be observed, see figures (\ref{fig:resonance_g}) and (\ref{fig:plot4a}), which indicates the existence of higher order Fourier coefficients in the electronic site's occupation. 

In addition to the resonance at $\Omega\approx\omega_c$, a higher, smaller resonance was observed at $\Omega\approx2\omega_c$, see figure (\ref{fig:resonance_e}), (\ref{fig:resonance_f}), (\ref{fig:resonance_g}) and (\ref{fig:resonance_h}). In contrast to resonance at $\Omega\approx\omega_c$, the resonance at $\Omega\approx2\omega_c$ is due to increases in the variance of phonon's position and momentum, which is calculated with the phonon lesser GF. This can be seen in figure (\ref{fig:resonance_h}), where the contributions from equation (\ref{eq:phonon_occupation}) are separated into the contributions from the mean position and momenta, $\left\langle Q_{\alpha} (t) \right\rangle^2 + \left\langle P_{\alpha} (t)\right\rangle^2$, and the variance terms, $i D^<_{\alpha\alpha} \left(t,t\right) + i D^{<,PP}_{\alpha\alpha} \left(t,t\right)$.

Within the parameter ranges investigated, the time-resolved observables were found be explained primarily by the first and second order Fourier coefficients. For the resonances at $\Omega \approx \omega_c$ and $\Omega \approx 2\omega_c$, the first order Fourier coefficient was the prominent contributor to the time-resolved dynamics. For the subharmonic resonance at $2\Omega \approx \omega_c$, the second order Fourier component was found to contribute significantly. This is expected given that this resonance is sensitive to the second order Fourier components within the electronic occupation, as seen in equations (\ref{eq:single_level_position}) and (\ref{eq:single_level_dr}).

Figure (\ref{fig:plot4}) also shows how the simplistic model fails to capture the resonance effects, whilst still capturing the general trend of the phonon occupation. This is understandable given the simplistic models disregards the driving's effects on central system's dynamics.

\section{Conclusion}

In this work we have investigated the periodic driving of a quantum dot with a Floquet nonequilibrium Green's function approach and $GD$ approximation. Specifically, the case of sinusoidal driving of the left lead was investigated for a single level and phonon system.

Particularly interesting for the stability of such driven systems, it was found that driving the lead energies in resonance with the vibrational frequency resulted in increased variations in average position, average momentum and occupation of the phonon mode. Moreover, whilst the time-averaged phonon occupation shows an increase in occupation when resonance occurs (see figure (\ref{fig:resonance_h}) and (\ref{fig:plot4})), the time-resolved result (figure (\ref{fig:resonance_g})) reveals more pronounced increases in occupation over the period of driving, reflecting the need to analyse time-resolved results when dealing to periodically driven systems.

Also discussed was a simple, phenomenological model that was found to replicate the time-averaged observables rather well in regimes away from resonance, particularly when the driving frequency was smaller than the vibrational frequency, see figure (\ref{fig:plot2}).

The method presented can be extended to many levels, with the addition of extra sites allowing for investigation of models where phonon modes may couple to many electronic sites or to the coupling between sites \cite{Park2011}. Furthermore, the effects of different waveforms for the driving could allow for novel means of probing and controlling junction dynamics. 

Adding full-counting statistic to the method could allow for the investigation of higher cumulants in the current, including zero-frequency noise \cite{Park2011,Honeychurch2020}. This could help answer and motivate questions surrounding the noisiness of signals passed through vibrationally active junctions, an important line of inquiry for functional devices. Additionally, statistics surrounding the phononic occupancies may be of importance, with the average occupations not being enough to evaluate the risk of device failure, given the possibility of large variances in occupation.

This study has made use of self-consistent perturbation theory, allowing small electron-vibrational coupling strengths relative to what is present in many junction architectures. Whether the findings of the investigation follow in stronger settings remains to be seen. However, one can hypothesise, that given an increase in electron-vibrational coupling results in more pronounced resonance effects in the present work, the effects of resonance would only become larger in methods able to handle far stronger couplings.  

\bibliographystyle{plain}
\bibliography{paper3}

\begin{thebibliography}{10}

\bibitem{Arielly2017}
Rani Arielly, Nirit Nachman, Yaroslav Zelinskyy, Volkhard May, and Yoram
  Selzer.
\newblock Picosecond time resolved conductance measurements of redox molecular
  junctions.
\newblock {\em The Journal of Chemical Physics}, 146(9):092306, 2017.

\bibitem{Thoss2022}
Jakob B\"atge, Amikam Levy, Wenjie Dou, and Michael Thoss.
\newblock Nonadiabatically driven open quantum systems under out-of-equilibrium
  conditions: Effect of electron-phonon interaction.
\newblock {\em Phys. Rev. B}, 106:075419, Aug 2022.

\bibitem{Beltako2019}
K.~Beltako, N.~Cavassilas, M.~Lannoo, and F.~Michelini.
\newblock Insights into the charge separation dynamics in photoexcited
  molecular junctions.
\newblock {\em The Journal of Physical Chemistry C}, 123(51):30885--30892,
  2019.

\bibitem{Bode2012}
Niels Bode, Silvia~Viola Kusminskiy, Reinhold Egger, and Felix von Oppen.
\newblock Current-induced forces in mesoscopic systems: A scattering-matrix
  approach.
\newblock {\em Beilstein Journal of Nanotechnology}, 3:144--162, 2012.

\bibitem{Brandes1997}
Tobias Brandes.
\newblock Truncation method for green's functions in time-dependent fields.
\newblock {\em Phys. Rev. B}, 56:1213--1224, Jul 1997.

\bibitem{Cabra2020}
Gabriel Cabra, Ignacio Franco, and Michael Galperin.
\newblock Optical properties of periodically driven open nonequilibrium quantum
  systems.
\newblock {\em The Journal of Chemical Physics}, 152(9):094101, 2020.

\bibitem{cuevas2010molecular}
J~C Cuevas and E~Scheer.
\newblock {\em Molecular Electronics: An Introduction to Theory and
  Experiment}.
\newblock EBSCO ebook academic collection. World Scientific Publishing Company
  Pte Limited, 2010.

\bibitem{Dash2010}
L.~K. Dash, H.~Ness, and R.~W. Godby.
\newblock Nonequilibrium electronic structure of interacting single-molecule
  nanojunctions: Vertex corrections and polarization effects for the
  electron-vibron coupling.
\newblock {\em The Journal of Chemical Physics}, 132(10):104113, 2010.

\bibitem{djukic2005stretching}
D~Djukic, Kristian~Sommer Thygesen, Carlos Untiedt, RHM Smit, Karsten~Wedel
  Jacobsen, and JM~Van~Ruitenbeek.
\newblock Stretching dependence of the vibration modes of a single-molecule pt-
  h 2- pt bridge.
\newblock {\em Physical Review B}, 71(16):161402, 2005.

\bibitem{Erpenbeck2019}
A.~Erpenbeck, L.~G{\"o}tzend{\"o}rfer, C.~Schinabeck, and M.~Thoss.
\newblock Hierarchical quantum master equation approach to charge transport in
  molecular junctions with time-dependent molecule-lead coupling strengths.
\newblock {\em The European Physical Journal Special Topics},
  227(15):1981--1994, Mar 2019.

\bibitem{Galperin2006}
Michael Galperin, Abraham Nitzan, and Mark~A. Ratner.
\newblock Resonant inelastic tunneling in molecular junctions.
\newblock {\em Phys. Rev. B}, 73:045314, Jan 2006.

\bibitem{Galperin2004a}
Michael Galperin, Mark~A. Ratner, and Abraham Nitzan.
\newblock Inelastic electron tunneling spectroscopy in molecular junctions:
  Peaks and dips.
\newblock {\em The Journal of Chemical Physics}, 121(23):11965--11979, 2004.

\bibitem{Galperin2004b}
Michael Galperin, Mark~A. Ratner, and Abraham Nitzan.
\newblock On the line widths of vibrational features in inelastic electron
  tunneling spectroscopy.
\newblock {\em Nano Letters}, 4(9):1605--1611, 2004.

\bibitem{Galperin2007}
Michael Galperin, Mark~A Ratner, and Abraham Nitzan.
\newblock Molecular transport junctions: vibrational effects.
\newblock {\em Journal of Physics: Condensed Matter}, 19(10):103201, feb 2007.

\bibitem{Haug2008}
H~Haug and A~P Jauho.
\newblock {\em {Quantum Kinetics in Transport and Optics of Semiconductors}},
  volume 123 of {\em Solid-State Sciences}.
\newblock Springer Berlin Heidelberg, Berlin, Heidelberg, 2008.

\bibitem{Haughian2016}
Patrick Haughian, Stefan Walter, Andreas Nunnenkamp, and Thomas~L. Schmidt.
\newblock Lifting the franck-condon blockade in driven quantum dots.
\newblock {\em Phys. Rev. B}, 94:205412, Nov 2016.

\bibitem{Haughian2017}
Patrick Haughian, Han~Hoe Yap, Jiangbin Gong, and Thomas~L. Schmidt.
\newblock Charge pumping in strongly coupled molecular quantum dots.
\newblock {\em Phys. Rev. B}, 96:195432, Nov 2017.

\bibitem{Honeychurch2019}
Thomas~D. Honeychurch and Daniel~S. Kosov.
\newblock Timescale separation solution of the kadanoff-baym equations for
  quantum transport in time-dependent fields.
\newblock {\em Phys. Rev. B}, 100:245423, Dec 2019.

\bibitem{Honeychurch2020}
Thomas~D. Honeychurch and Daniel~S. Kosov.
\newblock Full counting statistics for electron transport in periodically
  driven quantum dots.
\newblock {\em Phys. Rev. B}, 102:195409, Nov 2020.

\bibitem{Jauho1994}
Antti-Pekka Jauho, Ned~S. Wingreen, and Yigal Meir.
\newblock Time-dependent transport in interacting and noninteracting
  resonant-tunneling systems.
\newblock {\em Phys. Rev. B}, 50:5528--5544, Aug 1994.

\bibitem{Karlsson2020}
Daniel Karlsson and Robert van Leeuwen.
\newblock {\em Non-equilibrium Green's Functions for Coupled Fermion-Boson
  Systems}, pages 367--395.
\newblock Springer International Publishing, Cham, 2020.

\bibitem{Thoss2021}
Yaling Ke, André Erpenbeck, Uri Peskin, and Michael Thoss.
\newblock Unraveling current-induced dissociation mechanisms in single-molecule
  junctions.
\newblock {\em The Journal of Chemical Physics}, 154(23):234702, 2021.

\bibitem{Kershaw2020}
Vincent~F. Kershaw and Daniel~S. Kosov.
\newblock Non-adiabatic effects of nuclear motion in quantum transport of
  electrons: A self-consistent keldysh–langevin study.
\newblock {\em The Journal of Chemical Physics}, 153(15):154101, 2020.

\bibitem{Peskin2020}
Maayan Kuperman, Linoy Nagar, and Uri Peskin.
\newblock Mechanical stabilization of nanoscale conductors by plasmon
  oscillations.
\newblock {\em Nano Letters}, 20(7):5531--5537, 2020.
\newblock PMID: 32538634.

\bibitem{Lu2012}
Jing-Tao L\"u, Mads Brandbyge, Per Hedeg\aa{}rd, Tchavdar~N. Todorov, and
  Daniel Dundas.
\newblock Current-induced atomic dynamics, instabilities, and raman signals:
  Quasiclassical langevin equation approach.
\newblock {\em Phys. Rev. B}, 85:245444, Jun 2012.

\bibitem{Hedegard2010}
Jing-Tao Lü, Mads Brandbyge, and Per Hedegård.
\newblock Blowing the fuse: Berry’s phase and runaway vibrations in molecular
  conductors.
\newblock {\em Nano Letters}, 10(5):1657--1663, 2010.

\bibitem{Maier2011}
S.~Maier, T.~L. Schmidt, and A.~Komnik.
\newblock Charge transfer statistics of a molecular quantum dot with strong
  electron-phonon interaction.
\newblock {\em Phys. Rev. B}, 83:085401, Feb 2011.

\bibitem{Ochoa2015}
Maicol~A. Ochoa, Yoram Selzer, Uri Peskin, and Michael Galperin.
\newblock Pump–probe noise spectroscopy of molecular junctions.
\newblock {\em The Journal of Physical Chemistry Letters}, 6(3):470--476, 2015.
\newblock PMID: 26261965.

\bibitem{Park2011}
Tae-Ho Park and Michael Galperin.
\newblock Self-consistent full counting statistics of inelastic transport.
\newblock {\em Phys. Rev. B}, 84:205450, Nov 2011.

\bibitem{Darwish2020}
Chandramalika~R. Peiris, Simone Ciampi, Essam~M. Dief, Jinyang Zhang, Peter~J.
  Canfield, Anton~P. Le~Brun, Daniel~S. Kosov, Jeffrey~R. Reimers, and Nadim
  Darwish.
\newblock Spontaneous s–si bonding of alkanethiols to si(111)–h: towards
  si–molecule–si circuits.
\newblock {\em Chem. Sci.}, 11:5246--5256, 2020.

\bibitem{Peskin2017}
Uri Peskin.
\newblock Formulation of charge transport in molecular junctions with
  time-dependent molecule-leads coupling operators.
\newblock {\em Fortschritte der Physik}, 65(6-8):1600048, 2017.

\bibitem{Platero2004}
Gloria Platero and Ramón Aguado.
\newblock Photon-assisted transport in semiconductor nanostructures.
\newblock {\em Physics Reports}, 395(1):1 -- 157, 2004.

\bibitem{Preston2021_first_passage}
Riley~J. Preston, Maxim~F. Gelin, and Daniel~S. Kosov.
\newblock First-passage time theory of activated rate chemical processes in
  electronic molecular junctions.
\newblock {\em The Journal of Chemical Physics}, 154(11):114108, 2021.

\bibitem{Preston2020}
Riley~J. Preston, Thomas~D. Honeychurch, and Daniel~S. Kosov.
\newblock Cooling molecular electronic junctions by ac current.
\newblock {\em The Journal of Chemical Physics}, 153(12):121102, 2020.

\bibitem{preston2020current}
Riley~J Preston, Vincent~F Kershaw, and Daniel~S Kosov.
\newblock Current-induced atomic motion, structural instabilities, and negative
  temperatures on molecule-electrode interfaces in electronic junctions.
\newblock {\em Physical Review B}, 101(15):155415, 2020.

\bibitem{Rammer2007}
Jørgen Rammer.
\newblock {\em Quantum Field Theory of Non-equilibrium States}.
\newblock Cambridge University Press, 2007.

\bibitem{Ridley2018}
Michael Ridley, Viveka~N. Singh, Emanuel Gull, and Guy Cohen.
\newblock Numerically exact full counting statistics of the nonequilibrium
  anderson impurity model.
\newblock {\em Phys. Rev. B}, 97:115109, Mar 2018.

\bibitem{Ridley2022}
Michael Ridley, N.~Walter Talarico, Daniel Karlsson, Nicola Lo~Gullo, and Riku
  Tuovinen.
\newblock A many-body approach to transport in quantum systems: From the
  transient regime to the stationary state.
\newblock {\em Journal of Physics A: Mathematical and Theoretical}, 2022.

\bibitem{ryndyk2016theory}
Dmitry~A Ryndyk et~al.
\newblock Theory of quantum transport at nanoscale.
\newblock {\em Springer Series in Solid-State Sciences}, 184, 2016.

\bibitem{Schinabeck2020}
C.~Schinabeck and M.~Thoss.
\newblock Hierarchical quantum master equation approach to current fluctuations
  in nonequilibrium charge transport through nanosystems.
\newblock {\em Phys. Rev. B}, 101:075422, Feb 2020.

\bibitem{Schuler2016}
M.~Sch\"uler, J.~Berakdar, and Y.~Pavlyukh.
\newblock Time-dependent many-body treatment of electron-boson dynamics:
  Application to plasmon-accompanied photoemission.
\newblock {\em Phys. Rev. B}, 93:054303, Feb 2016.

\bibitem{Souto2015}
R.~Seoane~Souto, R.~Avriller, R.~C. Monreal, A.~Mart\'{\i}n-Rodero, and
  A.~Levy~Yeyati.
\newblock Transient dynamics and waiting time distribution of molecular
  junctions in the polaronic regime.
\newblock {\em Phys. Rev. B}, 92:125435, Sep 2015.

\bibitem{Souto2014}
R.~Seoane~Souto, A.~Levy Yeyati, A.~Mart\'{\i}n-Rodero, and R.~C. Monreal.
\newblock Dressed tunneling approximation for electronic transport through
  molecular transistors.
\newblock {\em Phys. Rev. B}, 89:085412, Feb 2014.

\bibitem{Souto2018}
R~Seoane Souto, R~Avriller, A~Levy Yeyati, and A~Mart{\'{\i}}n-Rodero.
\newblock Transient dynamics in interacting nanojunctions within
  self-consistent perturbation theory.
\newblock {\em New Journal of Physics}, 20(8):083039, aug 2018.

\bibitem{Stefanucci2013}
Gianluca Stefanucci and Robert van Leeuwen.
\newblock {\em Nonequilibrium Many-Body Theory of Quantum Systems: A Modern
  Introduction}.
\newblock Cambridge University Press, 2013.

\bibitem{Sakkinen2015}
Niko Säkkinen, Yang Peng, Heiko Appel, and Robert van Leeuwen.
\newblock Many-body green’s function theory for electron-phonon interactions:
  Ground state properties of the holstein dimer.
\newblock {\em The Journal of Chemical Physics}, 143(23):234101, 2015.

\bibitem{tal2008electron}
Oren Tal, M~Krieger, B~Leerink, and JM~Van~Ruitenbeek.
\newblock Electron-vibration interaction in single-molecule junctions: From
  contact to tunneling regimes.
\newblock {\em Physical review letters}, 100(19):196804, 2008.

\bibitem{Trasobares2016}
J.~Trasobares, D.~Vuillaume, D.~Th{\'e}ron, and N.~Cl{\'e}ment.
\newblock A 17{\thinspace}ghz molecular rectifier.
\newblock {\em Nature Communications}, 7(1):12850, Oct 2016.

\bibitem{Tu2006}
X.~W. Tu, J.~H. Lee, and W.~Ho.
\newblock Atomic-scale rectification at microwave frequency.
\newblock {\em The Journal of Chemical Physics}, 124(2):021105, 2006.

\bibitem{Ueda2011}
A.~Ueda, O.~Entin-Wohlman, and A.~Aharony.
\newblock Effects of coupling to vibrational modes on the ac conductance of
  molecular junctions.
\newblock {\em Phys. Rev. B}, 83:155438, Apr 2011.

\bibitem{Ueda2017}
A.~Ueda, Y.~Utsumi, Y.~Tokura, O.~Entin-Wohlman, and A.~Aharony.
\newblock Ac transport and full-counting statistics of molecular junctions in
  the weak electron-vibration coupling regime.
\newblock {\em J. Chem. Phys.}, 146(9):092313, 2017.

\bibitem{Ueda2016}
Akiko Ueda, Yasuhiro Utsumi, Hiroshi Imamura, and Yasuhiro Tokura.
\newblock Phonon-induced electron–hole excitation and ac conductance in
  molecular junction.
\newblock {\em Journal of the Physical Society of Japan}, 85(4):043703, 2016.

\bibitem{you2017recent}
Sifan You, Jing-Tao L{\"u}, Jing Guo, and Ying Jiang.
\newblock Recent advances in inelastic electron tunneling spectroscopy.
\newblock {\em Advances in Physics: X}, 2(3):907--936, 2017.

\end{thebibliography}

\end{document}